\documentstyle{mn}

\newif\ifAMStwofonts

\def\PsfigVersion{1.9}
\ifx\undefined\psfig\else \fi

\let\LaTeXAtSign=\@
\let\@=\relax
\edef\psfigRestoreAt{\catcode`\@=\number\catcode`@\relax}
\catcode`\@=11\relax
\newwrite\@unused
\def\ps@typeout#1{{\let\protect\string\immediate\write\@unused{#1}}}
\ps@typeout{psfig/tex \PsfigVersion}

\def\figurepath{./}

\def\@nnil{\@nil}
\def\@empty{}
\def\@psdonoop#1\@@#2#3{}
\def\@psdo#1:=#2\do#3{\edef\@psdotmp{#2}\ifx\@psdotmp\@empty \else
    \expandafter\@psdoloop#2,\@nil,\@nil\@@#1{#3}\fi}
\def\@psdoloop#1,#2,#3\@@#4#5{\def#4{#1}\ifx #4\@nnil \else
       #5\def#4{#2}\ifx #4\@nnil \else#5\@ipsdoloop #3\@@#4{#5}\fi\fi}
\def\@ipsdoloop#1,#2\@@#3#4{\def#3{#1}\ifx #3\@nnil 
       \let\@nextwhile=\@psdonoop \else
      #4\relax\let\@nextwhile=\@ipsdoloop\fi\@nextwhile#2\@@#3{#4}}
\def\@tpsdo#1:=#2\do#3{\xdef\@psdotmp{#2}\ifx\@psdotmp\@empty \else
    \@tpsdoloop#2\@nil\@nil\@@#1{#3}\fi}
\def\@tpsdoloop#1#2\@@#3#4{\def#3{#1}\ifx #3\@nnil 
       \let\@nextwhile=\@psdonoop \else
      #4\relax\let\@nextwhile=\@tpsdoloop\fi\@nextwhile#2\@@#3{#4}}
\ifx\undefined\fbox
\newdimen\fboxrule
\newdimen\fboxsep
\newdimen\ps@tempdima
\newbox\ps@tempboxa
\long\def\fbox#1{\leavevmode\setbox\ps@tempboxa\hbox{#1}\ps@tempdima\fboxrule
    \advance\ps@tempdima \fboxsep \advance\ps@tempdima \dp\ps@tempboxa
   \hbox{\lower \ps@tempdima\hbox
  {\vbox{\hrule height \fboxrule
          \hbox{\vrule width \fboxrule \hskip\fboxsep
          \vbox{\vskip\fboxsep \box\ps@tempboxa\vskip\fboxsep}\hskip 
                 \fboxsep\vrule width \fboxrule}
                 \hrule height \fboxrule}}}}
\fi
\newread\ps@stream
\newif\ifnot@eof       
\newif\if@noisy        
\newif\if@atend        
\newif\if@psfile       
{\catcode`\%=12\global\gdef\epsf@start{
\def\epsf@PS{PS}
\def\epsf@getbb#1{%
\openin\ps@stream=#1
\ifeof\ps@stream\ps@typeout{Error, File #1 not found}\else
   {\not@eoftrue \chardef\other=12
    \def\do##1{\catcode`##1=\other}\dospecials \catcode`\ =10
    \loop
       \if@psfile
	  \read\ps@stream to \epsf@fileline
       \else{
	  \obeyspaces
          \read\ps@stream to \epsf@tmp\global\let\epsf@fileline\epsf@tmp}
       \fi
       \ifeof\ps@stream\not@eoffalse\else
       \if@psfile\else
       \expandafter\epsf@test\epsf@fileline:. \\%
       \fi
          \expandafter\epsf@aux\epsf@fileline:. \\%
       \fi
   \ifnot@eof\repeat
   }\closein\ps@stream\fi}%
\long\def\epsf@test#1#2#3:#4\\{\def\epsf@testit{#1#2}
			\ifx\epsf@testit\epsf@start\else
\ps@typeout{Warning! File does not start with `\epsf@start'.  It may not be a PostScript file.}
			\fi
			\@psfiletrue} 
{\catcode`\%=12\global\let\epsf@percent=
\long\def\epsf@aux#1#2:#3\\{\ifx#1\epsf@percent
   \def\epsf@testit{#2}\ifx\epsf@testit\epsf@bblit
	\@atendfalse
        \epsf@atend #3 . \\%
	\if@atend	
	   \if@verbose{
		\ps@typeout{psfig: found `(atend)'; continuing search}
	   }\fi
        \else
        \epsf@grab #3 . . . \\%
        \not@eoffalse
        \global\no@bbfalse
        \fi
   \fi\fi}%
\def\epsf@grab #1 #2 #3 #4 #5\\{%
   \global\def\epsf@llx{#1}\ifx\epsf@llx\empty
      \epsf@grab #2 #3 #4 #5 .\\\else
   \global\def\epsf@lly{#2}%
   \global\def\epsf@urx{#3}\global\def\epsf@ury{#4}\fi}%
\def\epsf@atendlit{(atend)} 
\def\epsf@atend #1 #2 #3\\{%
   \def\epsf@tmp{#1}\ifx\epsf@tmp\empty
      \epsf@atend #2 #3 .\\\else
   \ifx\epsf@tmp\epsf@atendlit\@atendtrue\fi\fi}

\chardef\psletter = 11 
\chardef\other = 12

\newif \ifdebug 
\newif\ifc@mpute 
\c@mputetrue 

\let\then = \relax
\def\r@dian{pt }
\let\r@dians = \r@dian
\let\dimensionless@nit = \r@dian
\let\dimensionless@nits = \dimensionless@nit
\def\internal@nit{sp }
\let\internal@nits = \internal@nit
\newif\ifstillc@nverging
\def \Mess@ge #1{\ifdebug \then \message {#1} \fi}

{ 
	\catcode `\@ = \psletter
	\gdef \nodimen {\expandafter \n@dimen \the \dimen}
	\gdef \term #1 #2 #3%
	       {\edef \t@ {\the #1}
		\edef \t@@ {\expandafter \n@dimen \the #2\r@dian}%
		\t@rm {\t@} {\t@@} {#3}%
	       }
	\gdef \t@rm #1 #2 #3%
	       {{%
		\count 0 = 0
		\dimen 0 = 1 \dimensionless@nit
		\dimen 2 = #2\relax
		\Mess@ge {Calculating term #1 of \nodimen 2}%
		\loop
		\ifnum	\count 0 < #1
		\then	\advance \count 0 by 1
			\Mess@ge {Iteration \the \count 0 \space}%
			\Multiply \dimen 0 by {\dimen 2}%
			\Mess@ge {After multiplication, term = \nodimen 0}%
			\Divide \dimen 0 by {\count 0}%
			\Mess@ge {After division, term = \nodimen 0}%
		\repeat
		\Mess@ge {Final value for term #1 of 
				\nodimen 2 \space is \nodimen 0}%
		\xdef \Term {#3 = \nodimen 0 \r@dians}%
		\aftergroup \Term
	       }}
	\catcode `\p = \other
	\catcode `\t = \other
	\gdef \n@dimen #1pt{#1} 
}

\def \Divide #1by #2{\divide #1 by #2} 

\def \Multiply #1by #2
       {{
	\count 0 = #1\relax
	\count 2 = #2\relax
	\count 4 = 65536
	\Mess@ge {Before scaling, count 0 = \the \count 0 \space and
			count 2 = \the \count 2}%
	\ifnum	\count 0 > 32767 
	\then	\divide \count 0 by 4
		\divide \count 4 by 4
	\else	\ifnum	\count 0 < -32767
		\then	\divide \count 0 by 4
			\divide \count 4 by 4
		\else
		\fi
	\fi
	\ifnum	\count 2 > 32767 
	\then	\divide \count 2 by 4
		\divide \count 4 by 4
	\else	\ifnum	\count 2 < -32767
		\then	\divide \count 2 by 4
			\divide \count 4 by 4
		\else
		\fi
	\fi
	\multiply \count 0 by \count 2
	\divide \count 0 by \count 4
	\xdef \product {#1 = \the \count 0 \internal@nits}%
	\aftergroup \product
       }}

\def\r@duce{\ifdim\dimen0 > 90\r@dian \then   
		\multiply\dimen0 by -1
		\advance\dimen0 by 180\r@dian
		\r@duce
	    \else \ifdim\dimen0 < -90\r@dian \then  
		\advance\dimen0 by 360\r@dian
		\r@duce
		\fi
	    \fi}

\def\Sine#1%
       {{%
	\dimen 0 = #1 \r@dian
	\r@duce
	\ifdim\dimen0 = -90\r@dian \then
	   \dimen4 = -1\r@dian
	   \c@mputefalse
	\fi
	\ifdim\dimen0 = 90\r@dian \then
	   \dimen4 = 1\r@dian
	   \c@mputefalse
	\fi
	\ifdim\dimen0 = 0\r@dian \then
	   \dimen4 = 0\r@dian
	   \c@mputefalse
	\fi
	\ifc@mpute \then
		\divide\dimen0 by 180
		\dimen0=3.141592654\dimen0
		\dimen 2 = 3.1415926535897963\r@dian 
		\divide\dimen 2 by 2 
		\Mess@ge {Sin: calculating Sin of \nodimen 0}%
		\count 0 = 1 
		\dimen 2 = 1 \r@dian 
		\dimen 4 = 0 \r@dian 
		\loop
			\ifnum	\dimen 2 = 0 
			\then	\stillc@nvergingfalse 
			\else	\stillc@nvergingtrue
			\fi
			\ifstillc@nverging 
			\then	\term {\count 0} {\dimen 0} {\dimen 2}%
				\advance \count 0 by 2
				\count 2 = \count 0
				\divide \count 2 by 2
				\ifodd	\count 2 
				\then	\advance \dimen 4 by \dimen 2
				\else	\advance \dimen 4 by -\dimen 2
				\fi
		\repeat
	\fi		
			\xdef \sine {\nodimen 4}%
       }}

\def\Cosine#1{\ifx\sine\UnDefined\edef\Savesine{\relax}\else
		             \edef\Savesine{\sine}\fi
	{\dimen0=#1\r@dian\advance\dimen0 by 90\r@dian
	 \Sine{\nodimen 0}
	 \xdef\cosine{\sine}
	 \xdef\sine{\Savesine}}}	      

\def\psdraft{
	\def\@psdraft{0}
}
\def\psfull{
	\def\@psdraft{100}
}

\psfull

\newif\if@scalefirst
\def\psscalefirst{\@scalefirsttrue}
\def\psrotatefirst{\@scalefirstfalse}
\psrotatefirst

\newif\if@draftbox
\def\psnodraftbox{
	\@draftboxfalse
}
\def\psdraftbox{
	\@draftboxtrue
}
\@draftboxtrue

\newif\if@prologfile
\newif\if@postlogfile
\def\pssilent{
	\@noisyfalse
}
\def\psnoisy{
	\@noisytrue
}
\psnoisy
\newif\if@bbllx
\newif\if@bblly
\newif\if@bburx
\newif\if@bbury
\newif\if@height
\newif\if@width
\newif\if@rheight
\newif\if@rwidth
\newif\if@angle
\newif\if@clip
\newif\if@verbose
\newif\if@scale
\def\@p@@sclip#1{\@cliptrue}

\newif\if@decmpr

\def\@p@@sfigure#1{\def\@p@sfile{null}\def\@p@sbbfile{null}
	        \openin1=#1.bb
		\ifeof1\closein1
	        	\openin1=\figurepath#1.bb
			\ifeof1\closein1
			        \openin1=#1
				\ifeof1\closein1%
				       \openin1=\figurepath#1
					\ifeof1
					   \ps@typeout{Error, File #1 not found}
						\if@bbllx\if@bblly
				   		\if@bburx\if@bbury
			      				\def\@p@sfile{#1}%
			      				\def\@p@sbbfile{#1}%
							\@decmprfalse
				  	   	\fi\fi\fi\fi
					\else\closein1
				    		\def\@p@sfile{\figurepath#1}%
				    		\def\@p@sbbfile{\figurepath#1}%
						\@decmprfalse
	                       		\fi%
			 	\else\closein1%
					\def\@p@sfile{#1}
					\def\@p@sbbfile{#1}
					\@decmprfalse
			 	\fi
			\else
				\def\@p@sfile{\figurepath#1}
				\def\@p@sbbfile{\figurepath#1.bb}
				\@decmprtrue
			\fi
		\else
			\def\@p@sfile{#1}
			\def\@p@sbbfile{#1.bb}
			\@decmprtrue
		\fi}

\def\@p@@sfile#1{\@p@@sfigure{#1}}

\def\@p@@sbbllx#1{
		\@bbllxtrue
		\dimen100=#1
		\edef\@p@sbbllx{\number\dimen100}
}
\def\@p@@sbblly#1{
		\@bbllytrue
		\dimen100=#1
		\edef\@p@sbblly{\number\dimen100}
}
\def\@p@@sbburx#1{
		\@bburxtrue
		\dimen100=#1
		\edef\@p@sbburx{\number\dimen100}
}
\def\@p@@sbbury#1{
		\@bburytrue
		\dimen100=#1
		\edef\@p@sbbury{\number\dimen100}
}
\def\@p@@sheight#1{
		\@heighttrue
		\dimen100=#1
   		\edef\@p@sheight{\number\dimen100}
}
\def\@p@@swidth#1{
		\@widthtrue
		\dimen100=#1
		\edef\@p@swidth{\number\dimen100}
}
\def\@p@@srheight#1{
		\@rheighttrue
		\dimen100=#1
		\edef\@p@srheight{\number\dimen100}
}
\def\@p@@srwidth#1{
		\@rwidthtrue
		\dimen100=#1
		\edef\@p@srwidth{\number\dimen100}
}
\def\@p@@sangle#1{
		\@angletrue
		\edef\@p@sangle{#1} 
}
\def\@p@@srotate#1{\@p@@sangle{-#1}}
\def\@p@@sscale#1{
		\@scaletrue
		\edef\@p@sscale{#1}
}
\def\@p@@ssilent#1{ 
		\@verbosefalse
}
\def\@p@@sprolog#1{\@prologfiletrue\def\@prologfileval{#1}}
\def\@p@@spostlog#1{\@postlogfiletrue\def\@postlogfileval{#1}}
\def\@cs@name#1{\csname #1\endcsname}
\def\@setparms#1=#2,{\@cs@name{@p@@s#1}{#2}}
\def\ps@init@parms{
		\@bbllxfalse \@bbllyfalse
		\@bburxfalse \@bburyfalse
		\@heightfalse \@widthfalse
		\@rheightfalse \@rwidthfalse
		\@scalefalse
		\def\@p@sbbllx{}\def\@p@sbblly{}
		\def\@p@sbburx{}\def\@p@sbbury{}
		\def\@p@sheight{}\def\@p@swidth{}
		\def\@p@srheight{}\def\@p@srwidth{}
		\def\@p@sangle{0}
		\def\@p@sfile{} \def\@p@sbbfile{}
		\def\@p@scost{10}
		\def\@sc{}
		\@prologfilefalse
		\@postlogfilefalse
		\@clipfalse
		\if@noisy
			\@verbosetrue
		\else
			\@verbosefalse
		\fi
}
\def\parse@ps@parms#1{
	 	\@psdo\@psfiga:=#1\do
		   {\expandafter\@setparms\@psfiga,}}
\newif\ifno@bb
\def\bb@missing{
	\if@verbose{
		\ps@typeout{psfig: searching \@p@sbbfile \space  for bounding box}
	}\fi
	\no@bbtrue
	\epsf@getbb{\@p@sbbfile}
        \ifno@bb \else \bb@cull\epsf@llx\epsf@lly\epsf@urx\epsf@ury\fi
}	
\def\bb@cull#1#2#3#4{
	\dimen100=#1 bp\edef\@p@sbbllx{\number\dimen100}
	\dimen100=#2 bp\edef\@p@sbblly{\number\dimen100}
	\dimen100=#3 bp\edef\@p@sbburx{\number\dimen100}
	\dimen100=#4 bp\edef\@p@sbbury{\number\dimen100}
	\no@bbfalse
}
\newdimen\p@intvaluex
\newdimen\p@intvaluey
\def\rotate@#1#2{{\dimen0=#1 sp\dimen1=#2 sp
		  \global\p@intvaluex=\cosine\dimen0
		  \dimen3=\sine\dimen1
		  \global\advance\p@intvaluex by -\dimen3
		  \global\p@intvaluey=\sine\dimen0
		  \dimen3=\cosine\dimen1
		  \global\advance\p@intvaluey by \dimen3
		  }}
\def\compute@bb{
		\no@bbfalse
		\if@bbllx \else \no@bbtrue \fi
		\if@bblly \else \no@bbtrue \fi
		\if@bburx \else \no@bbtrue \fi
		\if@bbury \else \no@bbtrue \fi
		\ifno@bb \bb@missing \fi
		\ifno@bb \ps@typeout{FATAL ERROR: no bb supplied or found}
			\no-bb-error
		\fi
		\count203=\@p@sbburx
		\count204=\@p@sbbury
		\advance\count203 by -\@p@sbbllx
		\advance\count204 by -\@p@sbblly
		\edef\ps@bbw{\number\count203}
		\edef\ps@bbh{\number\count204}
		\if@angle 
			\Sine{\@p@sangle}\Cosine{\@p@sangle}
	        	{\dimen100=\maxdimen\xdef\r@p@sbbllx{\number\dimen100}
					    \xdef\r@p@sbblly{\number\dimen100}
			                    \xdef\r@p@sbburx{-\number\dimen100}
					    \xdef\r@p@sbbury{-\number\dimen100}}
                        \def\minmaxtest{
			   \ifnum\number\p@intvaluex<\r@p@sbbllx
			      \xdef\r@p@sbbllx{\number\p@intvaluex}\fi
			   \ifnum\number\p@intvaluex>\r@p@sbburx
			      \xdef\r@p@sbburx{\number\p@intvaluex}\fi
			   \ifnum\number\p@intvaluey<\r@p@sbblly
			      \xdef\r@p@sbblly{\number\p@intvaluey}\fi
			   \ifnum\number\p@intvaluey>\r@p@sbbury
			      \xdef\r@p@sbbury{\number\p@intvaluey}\fi
			   }
			\rotate@{\@p@sbbllx}{\@p@sbblly}
			\minmaxtest
			\rotate@{\@p@sbbllx}{\@p@sbbury}
			\minmaxtest
			\rotate@{\@p@sbburx}{\@p@sbblly}
			\minmaxtest
			\rotate@{\@p@sbburx}{\@p@sbbury}
			\minmaxtest
			\edef\@p@sbbllx{\r@p@sbbllx}\edef\@p@sbblly{\r@p@sbblly}
			\edef\@p@sbburx{\r@p@sbburx}\edef\@p@sbbury{\r@p@sbbury}
		\fi
		\count203=\@p@sbburx
		\count204=\@p@sbbury
		\advance\count203 by -\@p@sbbllx
		\advance\count204 by -\@p@sbblly
		\edef\@bbw{\number\count203}
		\edef\@bbh{\number\count204}
}
\def\in@hundreds#1#2#3{\count240=#2 \count241=#3
		     \count100=\count240	
		     \divide\count100 by \count241
		     \count101=\count100
		     \multiply\count101 by \count241
		     \advance\count240 by -\count101
		     \multiply\count240 by 10
		     \count101=\count240	
		     \divide\count101 by \count241
		     \count102=\count101
		     \multiply\count102 by \count241
		     \advance\count240 by -\count102
		     \multiply\count240 by 10
		     \count102=\count240	
		     \divide\count102 by \count241
		     \count200=#1\count205=0
		     \count201=\count200
			\multiply\count201 by \count100
		 	\advance\count205 by \count201
		     \count201=\count200
			\divide\count201 by 10
			\multiply\count201 by \count101
			\advance\count205 by \count201
		     \count201=\count200
			\divide\count201 by 100
			\multiply\count201 by \count102
			\advance\count205 by \count201
		     \edef\@result{\number\count205}
}
\def\ps@scaleinhundreds#1{
		\in@hundreds{#1}{\@p@sscale}{100}
		\edef#1{\@result}
}
\def\compute@wfromh{
		\in@hundreds{\@p@sheight}{\@bbw}{\@bbh}
		\edef\@p@swidth{\@result}
}
\def\compute@hfromw{
	        \in@hundreds{\@p@swidth}{\@bbh}{\@bbw}
		\edef\@p@sheight{\@result}
}
\def\compute@handw{
		\if@height 
			\if@width
			\else
				\compute@wfromh
			\fi
		\else 
			\if@width
				\compute@hfromw
			\else
				\edef\@p@sheight{\@bbh}
				\edef\@p@swidth{\@bbw}
			\fi
		\fi
}
\def\compute@resv{
		\if@rheight \else \edef\@p@srheight{\@p@sheight} \fi
		\if@rwidth \else \edef\@p@srwidth{\@p@swidth} \fi
}
\def\compute@sizes{
	\compute@bb
	\if@scalefirst\if@angle
	\if@width
	   \in@hundreds{\@p@swidth}{\@bbw}{\ps@bbw}
	   \edef\@p@swidth{\@result}
	\fi
	\if@height
	   \in@hundreds{\@p@sheight}{\@bbh}{\ps@bbh}
	   \edef\@p@sheight{\@result}
	\fi
	\fi\fi
	\compute@handw
	\compute@resv
	\if@scale
	   \if@verbose
	      \ps@typeout{(scaling by \@p@sscale)}%
	   \fi
	   \ps@scaleinhundreds{\@p@swidth}%
	   \ps@scaleinhundreds{\@p@sheight}%
	   \ps@scaleinhundreds{\@p@srwidth}%
	   \ps@scaleinhundreds{\@p@srheight}%
	\fi
}

\def\psfig#1{\vbox {
	\ps@init@parms
	\parse@ps@parms{#1}
	\compute@sizes
	\ifnum\@p@scost<\@psdraft{
		\special{ps::[begin] 	\@p@swidth \space \@p@sheight \space
				\@p@sbbllx \space \@p@sbblly \space
				\@p@sbburx \space \@p@sbbury \space
				startTexFig \space }
		\if@angle
			\special {ps:: \@p@sangle \space rotate \space} 
		\fi
		\if@clip{
			\if@verbose{
				\ps@typeout{(clip)}
			}\fi
			\special{ps:: doclip \space }
		}\fi
		\if@prologfile
		    \special{ps: plotfile \@prologfileval \space } \fi
		\if@decmpr{
			\if@verbose{
				\ps@typeout{psfig: including \@p@sfile.Z \space }
			}\fi
			\special{ps: plotfile "`zcat \@p@sfile.Z" \space }
		}\else{
			\if@verbose{
				\ps@typeout{psfig: including \@p@sfile \space }
			}\fi
			\special{ps: plotfile \@p@sfile \space }
		}\fi
		\if@postlogfile
		    \special{ps: plotfile \@postlogfileval \space } \fi
		\special{ps::[end] endTexFig \space }
		\vbox to \@p@srheight true sp{
			\hbox to \@p@srwidth true sp{
				\hss
			}
		\vss
		}
	}\else{
		\if@draftbox{		
			\hbox{\frame{\vbox to \@p@srheight true sp{
			\vss
			\hbox to \@p@srwidth true sp{ \hss \@p@sfile \hss }
			\vss
			}}}
		}\else{
			\vbox to \@p@srheight true sp{
			\vss
			\hbox to \@p@srwidth true sp{\hss}
			\vss
			}
		}\fi

	}\fi
}}
\psfigRestoreAt
\let\@=\LaTeXAtSign

\def\hh{$^h$}
\def\mm{$^m$}
\def\ss{$^s$}
\def\deg{\degr}
\def\asec{\arcsec}
\def\amin{\arcmin}
\def\etal{\rm et al.}
\def\msol{M$_{\sun}$}
\def\p{$\pm$}
\def\simlt{\lower.5ex\hbox{$\; \buildrel < \over \sim \;$}}
\def\simgt{\lower.5ex\hbox{$\; \buildrel > \over \sim \;$}}
\def\submitted{
}

\title[High-resolution Spectra of VLM Stars]
      {High-resolution Spectra of Very Low-Mass Stars\thanks{
Based on observations made at the European Southern Observatory 3.6-m telescope, La Silla, Chile,
and the 5-m Hale telescope, Palomar Observatory, USA.}}

\author[C.G. Tinney \& I.N. Reid]
       {C.G. Tinney$^{1,2}$ \& I.N. Reid$^{3}$ \\
     $^1$Anglo-Australian Observatory, PO Box 296, Epping. N.S.W. 2121. Australia. {\tt cgt@aaoepp.aao.gov.au}\\
     $^2$European Southern Observatory, Garching, Germany.\\
     $^3$Palomar Observatory, 105-24, California Institute of Technology, Pasadena, CA. 91125, USA.\\
       }

\date{Accepted ---.
      Received ---;
      in original form ---}

\pagerange{\pageref{firstpage}--\pageref{lastpage}}
\pubyear{1998}

\begin{document}

\maketitle

\label{firstpage}

\begin{abstract}
We present the results of high-resolution (1-0.4\,\AA) optical spectroscopy of a 
sample of very low-mass stars. 
These data are used to examine the kinematics
of the stars at the bottom of the hydrogen-burning main sequence. 
No evidence is found for a significant difference between the kinematics of 
the stars in our sample with I--K $>$ 3.5 (M$_{Bol} \ga 12.8$) and those
of more massive M-dwarfs (M$_{Bol} \approx 7-10$). 
A spectral atlas at high (0.4\,\AA) resolution for M8-M9+ stars is provided, and
the equivalent widths of Cs\,I, Rb\,I and H$\alpha$ lines present in our spectra are examined. 
We analyse our data to search for the presence of rapid rotation, and  find that the
brown dwarf LP\,944-20 is a member of the class of ``inactive, rapid rotators''. Such
objects seem to be common at and below the hydrogen burning main sequence. It seems
that in low-mass/low-temperature dwarf objects either the mechanism which heats
the chromosphere, or the mechanism which generates magnetic fields, is greatly suppressed.
\end{abstract}


\section{Introduction}

There have been suggestions in the recent literature of the existence of a young
($\la 10^8$yr) population of brown dwarfs in the solar neighbourhood
masquerading as VLM stars. Hawkins \& Bessell (1988) carried out
a survey using UK Schmidt telescope plates, and
claim that the reddest stars show a smaller (ie. younger) velocity dispersion than
their mid-M-type stars. Kirkpatrick \& McCarthy (1994) have shown that a {\sl linear}
extrapolation of their spectral-type versus mass relation implies
sub-stellar masses for spectral types later than M7.
In such a situation, there would be a strong bias towards detecting
only young, higher-luminosity (M$_{Bol} \la 13$) VLM dwarfs. 
More compelling, however, are the results of the CCD survey
of Kirkpatrick \etal\ (1994) which finds a significant excess of
M$_I > 12$ (M$_{Bol} \ga 12$) dwarfs in the southern Galactic hemisphere:
seven in 7.9 square degrees versus none within the 19.4 area surveyed 
at northern latitudes. Given that the Sun lies $\approx 30$pc above the
Galactic mid-plane, this would be consistent with the existence of a population of
very low scale-height M-dwarfs. Kirkpatrick \etal\ suggest that this population
is young either because such low-mass stars can only form at high metallicities,
{\em or} because the objects are substellar, and are detected only when young.
These hypotheses can be tested by estimating a kinematic age for 
local VLM dwarfs, and our current analysis centres on applying that test.
This work supercedes the preliminary
results presented by Reid, Tinney \& Mould (1994 - hereafter RTM).

\section{Observations \& Analysis}

\begin{table*}
  \caption{Radial Velocity Measurements -- Photometrically Selected Objects}
  \begin{tabular}{lrcrcccl}
Object              &\hfill V$_{hel}$\hfill\,&     UT         & V$_{mean}$       & I     & I--K &Phot.& Posn \\[5pt]
BRI\,0021-0214      &   +2.6 $\pm$ 5.0 & 20 Jun 1992 12:04:48 &     4.3$\pm$2.2  & 15.07 & 4.43 &3& 2  \\
                    &  +12.8 $\pm$ 4.1 & 21 Jun 1992 11:56:02 &                  &       &      & &    \\
                    &   -1.1 $\pm$ 6.5 & 28 Jul 1993 07:55:22 &                  &       &      & &    \\
                    &   +0.4 $\pm$ 3.5 & 29 Jul 1993 08:36:41 &                  &       &      & &    \\
TVLM\,831-161058    &  +27.0 $\pm$ 8.8 & 17 Oct 1992 09:36:55 &    21.2$\pm$5.1  & 16.63 & 3.96 &4& 1  \\
                    &  +32.0 $\pm$11.7 & 17 Oct 1992 10:12:42 &                  &       &      & &    \\
                    &  +13.0 $\pm$ 7.3 & 17 Oct 1992 10:40:46 &                  &       &      & &    \\
TVLM\,832-10443     &   +8.3 $\pm$ 6.2 & 17 Oct 1992 11:28:27 &   2.1$\pm$5.2$^c$& 16.06 & 4.07 &4& 1  \\ 
                    &  -13.6 $\pm$ 9.9 & 17 Oct 1992 12:05:06 &                  &       &      & &    \\
LP\,771-21/BR\,0246-1703$^{a,b}$
                    &  -25.6 $\pm$ 4.8 & 17 Oct 1992 08:50:09 & -25.6$\pm$4.8    & 15.42 & 3.95 &2&2,3 \\
LP\,944-20/BRI\,0337-3535$^b$
                    &   +4.6 $\pm$ 3.1 & 28 Jul 1993 09:53:58 &   7.4$\pm$1.3    & 14.16 & 4.58 &2&2,3 \\
                    &   +5.7 $\pm$ 2.1 & 29 Jul 1993 10:17:28 &                  &       &      & &    \\
                    &  +10.0 $\pm$ 2.0 & 09 Feb 1994 01:15:52 &                  &       &      & &    \\
BRI\,1222-1222      &   -0.6 $\pm$ 2.0 & 08 Feb 1994 09:00:17 &    -0.6$\pm$2.0  & 15.74 & 4.31 &2& 2,3\\
TVLM\,513-46546     &   +6.7 $\pm$ 4.2 & 19 Jun 1992 06:02:54 &     8.2$\pm$2.2  & 15.09 & 4.32 &4& 1  \\
                    &   +9.0 $\pm$ 3.6 & 20 Jun 1992 05:48:37 &                  &       &      & &    \\
                    &   +8.6 $\pm$ 3.7 & 20 Jun 1992 06:25:26 &                  &       &      & &    \\
TVLM\,513-42404A$^d$&  -42.0 $\pm$ 6.3 & 21 Jun 1992 07:43:14 & -40.9$\pm$4.6$^d$& 16.59 & 3.19 &4& 1  \\
                    &   -9.5 $\pm$ 6.5 & 21 Jun 1992 08:16:44 &                  &       &      & &    \\
                    &  -39.6 $\pm$ 6.8 & 21 Jun 1992 08:50:48 &                  &       &      & &    \\
TVLM\,513-8238      &  -16.8 $\pm$ 8.5 & 20 Jun 1992 07:56:28 &   -12.8$\pm$3.6  & 17.14 & 4.16 &4& 1  \\
                    &   -7.1 $\pm$ 7.2 & 20 Jun 1992 07:22:44 &                  &       &      & &    \\
                    &  -22.6 $\pm$ 7.8 & 21 Jun 1992 05:50:16 &                  &       &      & &    \\
                    &  -21.0 $\pm$ 9.9 & 21 Jun 1992 06:24:37 &                  &       &      & &    \\
                    &   -2.4 $\pm$ 7.3 & 21 Jun 1992 06:59:26 &                  &       &      & &    \\
TVLM\,868-110639    &  -47.3 $\pm$ 5.3 & 19 Jun 1992 06:52:07 &   -47.3$\pm$5.3  & 15.79 & 4.42 &4& 1  \\
TVLM\,890-60235     &  +14.8 $\pm$ 7.7 & 17 Oct 1992 04:42:59 &     2.6$\pm$3.8  & 16.65 & 3.55 &4& 1  \\
                    &   -3.3 $\pm$ 7.2 & 17 Oct 1992 05:20:07 &                  &       &      & &    \\
                    &   +6.0 $\pm$ 8.2 & 17 Oct 1992 05:58:31 &                  &       &      & &    \\
                    &   -5.5 $\pm$ 7.5 & 17 Oct 1992 06:32:37 &                  &       &      & &    \\
HB\,2124-4228       &   -5.0 $\pm$ 3.4 & 28 Jul 1993 05:34:31 &    -5.0$\pm$3.4  & 16.29 & 4.11 &2& 4  \\
BR\,2339-0447$^e$   &  +14.2 $\pm$ 2.1 & 28 Jul 1993 06:22:05 &    14.2$\pm$2.1  & 11.70 & 3.56 &3& 2  \\
  \end{tabular}
\raggedright
\noindent
\vskip 10pt
NOTES:\\
(a)\,Incorrectly referred to as BRI\,0246-1703 by Tinney, Mould \& Reid 1993.
(b)\,These objects, though originally catalogued as proper-motion objects, were 
independently selected as late stars by an optical photometric survey (Kirkpatrick, Henry \& Irwin 1997).
They are therefore included in the ``Photometric'' sample.
(c)\,A possible velocity variable, although the velocities are poor. 
(d)\,TVLM\,513-42404A has a common proper motion companion with magnitude I=18.56, I--K=4.35 (Tinney \etal\ 1995).
    The measurement of -9.5\,km/s was not used in calculating the mean.
(e)\,Classified as a giant (M7-8III) by Kirkpatrick, Henry \& Irwin 1997. Incorrectly referred to as BRI\,2339-0447
by Tinney, Mould \& Reid 1993.
(f)\,I--K photometry is not available. A spectral type of M5.5 has been assigned by Reid, Hawley \& Gizis 1995.
(g)\,Gl\,473 is a known binary with separation 0.2-0.8\asec\ and P$\approx$16\,yr (Henry \& McCarthy 1990). 
The components were not resolved by the spectrograph slit.
(h)\,VB\,8 was used as a velocity reference at ESO, so ESO observations are not included in the mean.
(i)\,VB\,10 was used as a velocity reference at Palomar, so Palomar observations are not included in the mean.
(j)\,Optical astrometry indicates that Gl\,831 may be an unresolved  binary with a period of under two years.
(Henry \& McCarthy 1993 and references therein).
(k)\,I--K photometry is not available. The spectral type is due to Kirkpatrick \etal\ 1991. In view
of the type mismatch between the object and template, this velocity should be treated with caution.
(l)\,GL\,866AB is a known triple system, with two of the components having $\sim 0.5\asec$ separation 
and P=2.20\,yr (Leinert \etal\ 1990).
This probably explains the observed velocity variations.  The components were not resolved by the 
spectrograph slit. The ``mean'' velocity shown should be interpreted with care in light of this.

PHOTOMETRY REFERENCES:\\
(1) Leggett 1992; (2) Tinney 1996; (3) Tinney, Mould \& Reid 1993; (4) Kirkpatrick, Henry \& Irwin 1997; (6) Hawkins \&
Bessell 1988; (7) Gullixson \etal\ 1995 and Harrington \etal\ 1993; (8) Weis 1988.

POSITION REFERENCES:\\
(1) Tinney 1993; (2) Tinney, Mould \& Reid 1993; (3) Kirkpatrick, Henry \& Irwin 1997; (4) Hawkins \& Bessell 1988; (5)
Gliese \& Jahreiss 1991; (6) Luyten 1979; (7) Tinney 1996.
\label{table_phot}
\end{table*}

\begin{table*}
  \caption{Radial Velocity Measurements -- Proper-motion Selected Objects}
  \begin{tabular}{lrcrccccl}
Object              &\hfill V$_{hel}$\hfill\,&     UT         & V$_{mean}$       & I     & I--K &Phot.& Posn \\[5pt]
\multicolumn{4}{l}{} \\
LHS\,2/GJ\,1002     &  -35.7 $\pm$ 5.5 & 17 Oct 1992 07:32:37 &   -35.7$\pm$5.5  & 10.16 & 2.74 &1& 5  \\
LHS\,112            &  +14.0 $\pm$ 4.0 & 17 Oct 1992 07:53:18 &    14.1$\pm$4.0  & 12.41 & M5.5$^f$ &8& 6  \\
ESO\,207-61         & +116.6 $\pm$ 2.1 & 08 Feb 1994 02:17:06 &   116.4$\pm$1.5  & 16.35 & 4.18 &2& 7  \\
                    & +116.3 $\pm$ 2.0 & 09 Feb 1994 03:04:46 &                  &       &      & &    \\
LHS\,234/Gl\,283B   &  -30.8 $\pm$ 2.0 & 09 Feb 1994 04:25:04 &   -30.8$\pm$2.0  & 12.43 & 3.17 &1& 6  \\
LHS\,248/GJ\,1111   &  +47.3 $\pm$ 3.2 & 17 Oct 1992 12:57:13 &    47.3$\pm$3.2  & 10.53 & 3.27 &1& 5  \\
LHS\,2065           &   +7.8 $\pm$ 4.5 & 17 Oct 1992 12:51:35 &     9.0$\pm$1.8  & 14.44 & 4.46 &1& 6  \\
                    &   +9.2 $\pm$ 2.0 & 08 Feb 1994 04:22:29 &                  &       &      & &    \\
LHS\,36/Gl\,406     &  +20.0 $\pm$ 3.3 & 21 Jun 1992 03:45:04 &    20.0$\pm$3.3  &  9.39 & 3.31 &1& 5  \\
LHS\,333/Gl\,473AB  &   +0.6 $\pm$ 3.3 & 19 Jun 1992 03:31:54 &   0.9$\pm$1.7$^g$&  8.92 & 2.86 &1& 5  \\
                    &   -2.4 $\pm$ 3.6 & 19 Jun 1992 03:43:00 &                  &       &      & &    \\
                    &   +3.6 $\pm$ 3.5 & 20 Jun 1992 03:33:03 &                  &       &      & &    \\
                    &   +1.8 $\pm$ 3.4 & 21 Jun 1992 03:25:15 &                  &       &      & &    \\
LHS\,2351           &   +2.3 $\pm$ 2.0 & 09 Feb 1994 05:44:56 &     2.3$\pm$2.0  & 14.91 & 3.57 &1& 6  \\
LHS\,2397a          &  +34.3 $\pm$ 2.1 & 08 Feb 1994 06:13:18 &    34.1$\pm$1.5  & 15.05 & 4.26 &2& 6  \\
                    &  +34.0 $\pm$ 2.1 & 09 Feb 1994 06:59:29 &                  &       &      & &    \\
LHS\,2875           &  -44.0 $\pm$ 2.1 & 29 Jul 1993 23:42:57 &   -45.5$\pm$1.4  & 10.58 & 2.05 &1& 6  \\
                    &  -46.9 $\pm$ 2.0 & 09 Feb 1994 07:59:21 &                  &       &      & &    \\
LHS\,2876           &  -48.6 $\pm$ 2.0 & 09 Feb 1994 08:40:30 &   -48.6$\pm$2.0  & 15.76 & 3.59 &2& 6  \\
LHS\,2924           &  -32.5 $\pm$ 4.5 & 19 Jun 1992 05:23:58 &   -33.4$\pm$2.9  & 15.21 & 4.54 &1& 6  \\
                    &  -34.0 $\pm$ 3.7 & 21 Jun 1992 05:06:08 &                  &       &      & &    \\
LHS\,2930           &  -10.7 $\pm$ 3.3 & 20 Jun 1992 04:31:55 &   -10.7$\pm$3.3  & 13.31 & 3.59 &1& 6  \\
LHS\,3003           &   +1.9 $\pm$ 2.0 & 09 Feb 1994 07:35:41 &     1.9$\pm$2.0  & 12.66 & 3.66 &2& 6  \\
LHS\,427/Gl\,643    &  +22.3 $\pm$ 5.7 & 17 Oct 1992 02:16:59 &    17.4$\pm$1.4  &  9.04 & 2.30 &1& 5  \\
                    &  +15.3 $\pm$ 2.0 & 28 Jul 1993 02:50:07 &                  &       &      & &    \\
                    &  +18.8 $\pm$ 2.0 & 08 Feb 1994 09:37:38 &                  &       &      & &    \\
LHS\,428/Gl\,644AB  &  +20.2 $\pm$ 6.2 & 17 Oct 1992 02:10:56 &    19.3$\pm$1.6  &  6.55 & 2.16 &1& 5  \\
                    &  +19.4 $\pm$ 3.2 & 28 Jul 1993 03:04:22 &                  &       &      & &    \\
                    &  +19.2 $\pm$ 2.0 & 08 Feb 1994 09:42:51 &                  &       &      & &    \\
LHS\,429/Gl\,644C/VB\,8 
                    &  +15.7 $\pm$ 3.7 & 17 Oct 1992 02:35:48 &  16.1$\pm$1.4$^h$& 12.24 & 3.42 &1& 5  \\
                    &  +16.4 $\pm$ 3.1 & 19 Jun 1992 06:57:54 &                  &       &      & &    \\
                    &  +15.2 $\pm$ 3.2 & 19 Jun 1992 08:38:31 &                  &       &      & &    \\
                    &  +17.5 $\pm$ 3.3 & 20 Jun 1992 06:33:39 &                  &       &      & &    \\
                    &  +15.8 $\pm$ 3.3 & 21 Jun 1992 09:01:32 &                  &       &      & &    \\
                    &  +13.3 $\pm$ 2.1 & 29 Jul 1993 02:24:31 &                  &       &      & &    \\
                    &  +14.4 $\pm$ 2.0 & 09 Feb 1994 09:31:43 &                  &       &      & &    \\
LHS\,57/Gl\,699     & -112.3 $\pm$ 5.1 & 19 Jun 1992 08:40:35 &  -112.3$\pm$5.1  &  6.77 & 2.25 &1& 5  \\
LHS\,474/Gl\,752B/VB\,10
                    &  +34.8 $\pm$ 3.1 & 20 Jun 1992 09:27:42 &  35.3$\pm$1.5$^i$& 12.80 & 3.98 &1& 5  \\
                    &  +34.9 $\pm$ 2.1 & 28 Jul 1993 03:32:50 &                  &       &      & &    \\                                      \\
                    &  +35.7 $\pm$ 2.1 & 29 Jul 1993 04:25:38 &                  &       &      & &    \\
LHS\,511/Gl\,831    &  -56.5 $\pm$ 2.0 & 28 Jul 1993 05:47:10 & -56.5$\pm$2.0$^j$&  9.02 & 2.60 &1& 5  \\
LHS\,515            &  +59.5 $\pm$ 7.7 & 17 Oct 1992 03:33:19 &    59.5$\pm$7.7  & 15.16 & sdM2.5$^k$& & 6  \\
LHS\,523            &   -7.4 $\pm$ 3.9 & 17 Oct 1992 03:48:55 &    -7.4$\pm$3.9  & 13.00 & 3.10 &1& 6  \\
LHS\,68/Gl\,866AB   &  -43.2 $\pm$ 3.7 & 19 Jun 1992 11:18:37 & -49.6$\pm$2.8$^j$&  8.62 & 3.06 &1& 5  \\
                    &  -58.2 $\pm$ 4.3 & 21 Jun 1992 11:54:01 &                  &       &      & &    \\
  \end{tabular}
\raggedright
\noindent
\vskip 10pt
NOTES, PHOTOMETRY and POSITION REFERENCES: See Table \ref{table_phot}.\\
\label{table_prop}
\end{table*}

While proper motions are straight-forward (if tedious) to acquire for even very
faint M-dwarfs, radial velocities are more difficult to measure. The low
intrinsic luminosities of late-type dwarfs lead to only a relatively
small number of those stars being accessible
to accurate velocity measurement, particularly since
those velocities should be obtained 
by cross-correlation on the stellar pseudo-continuum. 
The oft-used H$\alpha$ emission line is produced via chromospheric activity
and may not be a reliable velocity indicator. The present sample includes 
37 stars or systems bright enough in the 7000-9000\AA\ region (I$\la$17) to 
obtain radial velocities -- 20 of these have M$_{Bol}\geq$12.8.

\subsection{ESO 3.6-m}

Spectra were obtained using the Cassegrain Echelle Spectrograph
(CASPEC) on the ESO 3.6m telescope on the nights of 1993 July 28-29 
and 1994 February 8-9 (UT). 
The full width at half maximum (FWHM) resolution was 16\,km/s (eg. 0.43\AA\ at 8000\AA), at
a dispersion of 9\,km/s per pixel. The wavelength range covered was
6400\AA\ to 9100\AA, though redwards of 8045\AA\ the wavelength coverage is not
complete and suffers inter-order gaps. For the faintest stars observed at ESO (I$\la$16.3)
exposures of up to 4 hours were used. Despite the fact that the signal-to-noise
in an individual pixel at this resolution is low (S/N$\approx$2.5
at 6500\AA, S/N$\approx$30 at 9200\AA), the long wavelength coverage of
the instrument means that the final velocity precision is limited by the
flexure, rather than S/N.
Flexure is severe in CASPEC, so considerable care was taken
in the observing and data reduction. Arc spectra were taken before and after each exposure, in
order to provide a mean wavelength zero-point. The instrumental velocity system
so constructed was checked by cross-correlating the wavelength calibrated night-sky
spectra against each other  -- from which we derive a systematic uncertainty due to flexure 
of 2\,km/s.

The radial velocity system was calibrated initially using observations of the
preliminary IAU M-giant standard $\alpha$\,Cet (V$_{hel} = -25.8\pm0.5$\,km/s; Trans.IAU,XXIB,275).
By cross-correlation with $\alpha$\,Cet we determined the heliocentric radial 
velocity of VB\,8 to be 14.5$\pm$1.0\,km/s and of VB\,10 to be 34.4$\pm$1.3\,km/s. 
VB\,8 was then adopted as the cross-correlation template for {\em all of the CASPEC data}, 
with an assumed  V$_{hel} = 14.5$\,km/s. Cross-correlations were carried out on
an order-by-order basis, using regions of the spectrum free
of atmospheric absorption and strong sky lines. The results for each order were then
combined as a weighted mean to produce a velocity shift for each exposure.
Tables \ref{table_phot} and \ref{table_prop} show measured radial velocities for
each observation. The uncertainties 
quoted for individual observations are based on
the uncertainty produced by the cross-correlation procedure of Tonry \& Davis (1979)\footnote{
RTM used an empirical technique to estimate cross-correlation velocity
uncertainties, together with the expressions of Tonry \& Davis (1979). Based
on this it was stated that the Tonry \& Davis algorithm gave systematically
low uncertainties. It is now known that this was due to an error in our
implementation of their algorithm -- the correct algorithm gives similar
uncertainties to our empirical technique.}, combined with the
systematic 2\,km/s uncertainty due to spectrograph flexure -- for most
targets it is this systematic uncertainty which dominates.

\subsection{Palomar 5-m}

Radial velocities were obtained using the 
red arm of the Double Spectrograph on
the Hale 5m telescope, on 1992 June 19-22 (UT) and 1992 October 16-17 (UT). 
A 1200 line/mm grating was used in the following
configurations: 76\,km/s (2.15\AA) resolution at a dispersion of 28\,km/s (0.80\AA) per pixel, 
over 8110-8735\AA\ in June 1992; and 74\,km/s  (1.78\AA) resolution at
a dispersion of 36\,km/s (0.81\AA) per pixel over 7906--7549\AA.
These are the same observations 
reported by RTM. However, in the course of checking our CASPEC radial velocities 
against both the literature and the results of RTM, it became clear that
there were problems with the RTM velocities, which at that time could
not be checked against other observations owing to a lack of published
velocities for late M-dwarfs.

We have re-reduced the RTM data, taking the care to ensure a rigid
velocity system with the CASPEC data. Calibration arcs show 
flexure in the Double Spectrograph of up to 20\,km/s over the course of a night, 
so arcs were taken before and after each
exposure of longer then 15 minutes. 
Cross correlations of arcs before and after long exposures reveal
no flexure larger than 5\,km/s. We adopt a lower limit to the precision of
our velocities of 3\,km/s based on our ability to calibrate out this flexure.

VB\,10 was adopted as the velocity template,
with an assumed V$_{hel} = 35$\,km/s (based on our CASPEC observations). 
VB\,10 was adopted in preference to VB\,8, as it was felt that at the lower spectral
resolution of the Palomar data a good spectral-type match between template and object 
is more important. The results of this re-reduction are also shown in Tables \ref{table_phot} and \ref{table_prop}.

\subsection{Combined Results}

\begin{table}
  \caption{Radial velocities compared with other studies.}
  \begin{tabular}{lrrr}
Object                  &V${hel}$ & V${hel}$ & $\Delta$V$_{hel}$\\
                        & Literature & This work \\[10pt]
\multicolumn{4}{c}{Marcy \& Benitz (1989)}\\
Gl\,699                 & -110.9$\pm$0.2 & -112.3$\pm$5.1 &  1.4$\pm$5.1 \\
Gl\,831                 &  -57.1$\pm$1.0 &  -56.5$\pm$2.0 & -0.6$\pm$2.2 \\[10pt]
\multicolumn{4}{c}{Delfosse, Forveille, Perrier \& Mayor (1998)}\\
GJ\,1002                & -41$\pm$1 & -35.7$\pm$5.5 & -5.3$\pm$5.1\\
GJ\,1111                &   9$\pm$1 &  47.3$\pm$3.2 & -38$\pm$3.3\\
Gl\,406$^a$             &19.2$\pm$0.2&  20.0$\pm$3.3 & -0.8$\pm$3.3\\
Gl\,699                 & -111$\pm$1&-112.3$\pm$5.1 & -1.3$\pm$5.2\\[10pt]
\multicolumn{4}{c}{Basri \& Marcy (1995)}\\
GJ\,1111                &   11.4$\pm$3$^b$ &  47.3$\pm$3.2 & -36$\pm$4\\
BRI\,0021-0214          &   15.6$\pm$3$^b$ &   4.3$\pm$2.2 & 11.3$\pm$3.7\\[10pt]
  \end{tabular}
\raggedright
\noindent
\vskip 10pt
NOTES:\\
$a$ -- Gl\,406 velocity is not that from Table 2 of Delfosse \etal, but the more
precise value quoted in Mart\'{\i}n \etal\ 1997. $b$ -- Basri \& Marcy used Gl\,406
as a velocity standard with an assumed velocity of 16.5\,km/s. The velocities they
give have been corrected to using more precise velocity for Gl\,406 given by Delfosse \etal\ 1998.
\label{compare}
\end{table}

The final velocities are weighted means of the individual observations. 
Amongst the current sample, only Gl\,866AB shows compelling evidence for 
velocity variability.
The velocity difference seen in exposures separated by only 2 days is 15.0$\pm$5.7\,km/s.
Gl\,866AB has been resolved by infrared speckle techniques as a binary system with 
period of 2.2 years and separation of between 0.2-0.8\asec\ by Leinert \etal\ (1990).
However, Leinert \etal\ (1997) report a private communication from A.Duquennoy \& D.Latham
to the effect that a third component has been detected spectroscopically. Reid \& Gizis (1997)
also report the detection of an SB2 component, with a period inconsistent with
that of the resolved components. Our current results add further support for
the presence of a third component in this system. 

TVLM\,832-10443 and TVLM\,513-42404A both show possible evidence for velocity variability, 
although the data are of poor quality. In the case of TVLM\,513-42404A
where three observations are available, we have rejected the discrepant value 
to estimate a mean velocity. For TVLM\,832-10443 we use the mean of the two 
available velocities.

In Table \ref{compare} we compare the velocities derived in this study with those obtained by
other studies -- in particular those of Marcy \& Benitz (1989), Delfosse \etal\ (1998)
and Basri \& Marcy (1995). An examination of this table shows that with only two
exceptions our velocities are all consistent within 1-$\sigma$ uncertainties with
those obtained from the literature. 

Those exceptions are GJ\,1111 and BRI\,0021-0214. In the case of GJ\,1111 we can find no 
plausible explanation for the velocity difference observed. The remote possibility that
that GJ\,1111 is a radial velocity variable is argued against by the consistent velocities for it
determined at several epochs over the last 2 years at Observatoire de Haute Provence (X.Delfosse, priv.comm.). 
The target observed and the data reduction for this object have been throughly checked. We can find
no fault to which this discrepancy (which corresponds to approximately one pixel on our detector) 
can be adduced. The observations of LHS\,2065, VB\,8 and Gl\,643
made on the same night as GJ\,1111, show that the velocity system for this night was 
consistent with that obtained on all other observing runs.

\begin{table*}
  \caption{Derived Velocities}
  \begin{tabular}{lccccccc}
Object                   & V$_{tan}$            &     U$^a$      &        V$^a$   &       W$^a$    &  V$_{total}$$^b$ & Sample$^c$ & Astrometry\\
                         & km/s                 & km/s           & km/s           & km/s           &  km/s            &            & References\\[5pt]
BRI\,0021-0214           &	   8.5$\pm$ 0.5 &  -8.8$\pm$ 0.5 &  -3.2$\pm$ 1.5 &  -2.6$\pm$ 2.6 &   9.7$\pm$ 3.0 & A & 1\\
TVLM\,831-161058         &	  63.6$\pm$ 8.2 & -65.5$\pm$ 6.0 & -43.2$\pm$ 4.6 &  10.6$\pm$ 5.9 &  79.1$\pm$ 9.6 & A & 1\\
TVLM\,832-10443          &	  26.3$\pm$ 2.4 &   9.5$\pm$ 3.4 &  -5.4$\pm$ 2.8 & -23.9$\pm$ 3.6 &  26.3$\pm$ 5.7 & A & 1\\
LP\,771-21/BR\,0246-1703 &	  21.8$\pm$ 2.2 &  16.1$\pm$ 3.0 & -24.0$\pm$ 2.9 &  12.3$\pm$ 3.2 &  31.4$\pm$ 5.3 & A & 2\\
LP\,944-20/BRI\,0337-3535&	  10.3$\pm$ 0.3 & -20.6$\pm$ 0.8 & -15.9$\pm$ 0.8 &  -7.3$\pm$ 0.7 &  27.1$\pm$ 1.3 & A & 2\\
BRI\,1222-1222           &	  26.0$\pm$ 1.9 & -20.8$\pm$ 1.4 & -30.8$\pm$ 1.5 & -18.2$\pm$ 1.8 &  41.3$\pm$ 2.8 & A & 2\\
TVLM\,513-46546          &	   4.0$\pm$ 0.4 &  -4.0$\pm$ 1.3 & -12.5$\pm$ 1.2 &   1.5$\pm$ 1.4 &  13.2$\pm$ 2.2 & A & 1\\
TVLM\,513-42404A         &	  17.2$\pm$ 5.1 & -30.6$\pm$ 3.9 & -37.6$\pm$ 3.4 & -34.2$\pm$ 4.0 &  59.3$\pm$ 6.8 &   & 1\\
TVLM\,513-8238           &	  24.2$\pm$ 4.8 & -19.6$\pm$ 3.4 & -36.2$\pm$ 3.0 &  -6.5$\pm$ 3.5 &  41.7$\pm$ 6.0 & A & 1\\
TVLM\,868-110639         &	  31.4$\pm$ 2.6 & -60.3$\pm$ 3.7 & -28.7$\pm$ 2.8 & -22.8$\pm$ 3.6 &  70.5$\pm$ 5.9 & A & 1\\
TVLM\,890-60235          &	  17.7$\pm$ 2.0 &   7.9$\pm$ 2.1 &  -8.8$\pm$ 2.1 &  -0.5$\pm$ 3.1 &  11.8$\pm$ 4.3 & A & 1\\
HB\,2124-4228            &	  72.2$\pm$15.5 & -17.7$\pm$ 8.1 & -82.8$\pm$10.0 &  -6.5$\pm$ 9.1 &  84.9$\pm$15.8 & A & 1\\
LHS\,2/GJ\,1002          &	  45.5$\pm$ 0.8 &  27.9$\pm$ 0.7 & -49.6$\pm$ 2.8 &  16.1$\pm$ 4.8 &  59.2$\pm$ 5.6 &   & 2\\
LHS\,112                 &	  81.5$\pm$ 4.0 & -52.0$\pm$ 2.7 & -50.1$\pm$ 3.1 & -64.8$\pm$ 3.5 &  97.0$\pm$ 5.7 &   & 4\\
ESO\,207-61              &	  34.3$\pm$ 2.9 & -61.0$\pm$ 2.3 &-118.1$\pm$ 2.0 & -29.4$\pm$ 1.3 & 136.1$\pm$ 3.3 & B & 2\\
LHS\,234/Gl\,283B        &	  53.0$\pm$ 2.5 &  45.7$\pm$ 1.8 & -10.6$\pm$ 1.6 &  21.6$\pm$ 1.9 &  51.7$\pm$ 3.2 &   & 4\\
LHS\,248/GJ\,1111        &	  22.2$\pm$ 0.4 & -57.0$\pm$ 2.1 & -30.3$\pm$ 1.7 &   0.9$\pm$ 1.7 &  64.6$\pm$ 3.2 &   & 4\\
LHS\,2065                &	  22.0$\pm$ 0.3 & -23.1$\pm$ 1.2 & -22.1$\pm$ 0.9 & -21.6$\pm$ 1.1 &  38.6$\pm$ 1.8 & B & 3\\
LHS\,36/Gl\,406          &	  53.1$\pm$ 0.3 & -36.9$\pm$ 0.9 & -58.8$\pm$ 1.6 & -18.9$\pm$ 2.7 &  71.9$\pm$ 3.3 &   & 4\\
LHS\,333/Gl\,473AB       &	  37.4$\pm$ 0.7 & -43.0$\pm$ 0.7 & -26.4$\pm$ 0.9 &  -6.2$\pm$ 1.5 &  50.9$\pm$ 1.9 &   & 4\\
LHS\,2351                &	  49.0$\pm$ 4.3 & -54.8$\pm$ 2.9 &   6.2$\pm$ 2.9 &  -2.0$\pm$ 2.4 &  55.2$\pm$ 4.8 & B & 4\\
LHS\,2397a               &	  38.3$\pm$ 2.5 & -39.0$\pm$ 2.1 & -51.7$\pm$ 1.3 &   2.6$\pm$ 1.4 &  64.8$\pm$ 2.9 & B & 3\\
LHS\,2875                &	 102.6$\pm$12.6 &-116.5$\pm$ 9.0 & -41.9$\pm$ 6.3 &   2.3$\pm$ 6.3 & 123.9$\pm$12.7 &   & 2\\
LHS\,2876                &	 102.6$\pm$12.6 &-118.2$\pm$ 9.0 & -41.3$\pm$ 6.3 &  -0.2$\pm$ 6.4 & 125.2$\pm$12.8 & B & 2\\
LHS\,2924                &	  40.0$\pm$ 0.6 &  -1.8$\pm$ 1.5 & -57.5$\pm$ 1.7 & -28.4$\pm$ 1.9 &  64.2$\pm$ 3.0 & B & 3\\
LHS\,2930                &	  37.1$\pm$ 0.5 & -37.0$\pm$ 1.6 & -37.5$\pm$ 2.3 &  -4.4$\pm$ 1.8 &  52.9$\pm$ 3.3 & B & 3\\
LHS\,3003                &	  28.8$\pm$ 1.0 & -12.6$\pm$ 1.3 & -37.1$\pm$ 1.3 & -17.7$\pm$ 1.3 &  43.0$\pm$ 2.2 & B & 2\\
LHS\,427/Gl\,643         &	  32.7$\pm$ 1.3 &  11.6$\pm$ 1.3 & -39.8$\pm$ 1.0 &   4.8$\pm$ 0.9 &  41.7$\pm$ 1.9 &   & 4\\
LHS\,428/Gl\,644AB       &	  36.2$\pm$ 0.6 &  13.9$\pm$ 1.4 & -42.9$\pm$ 0.8 &   6.0$\pm$ 0.6 &  45.5$\pm$ 1.7 &   & 4\\
LHS\,429/Gl\,644C/VB\,8  &	  36.0$\pm$ 0.6 &   9.4$\pm$ 1.7 & -43.6$\pm$ 1.0 &   4.4$\pm$ 0.7 &  44.8$\pm$ 2.1 &   & 3\\
LHS\,57/Gl\,699          &	  89.4$\pm$ 0.2 &-151.3$\pm$ 4.4 &  -7.2$\pm$ 2.3 &  11.9$\pm$ 1.0 & 152.0$\pm$ 5.1 &   & 4\\
LHS\,474/Gl\,752B/VB\,10 &	  42.4$\pm$ 1.8 &  44.9$\pm$ 1.8 & -20.3$\pm$ 1.5 & -11.8$\pm$ 1.1 &  50.6$\pm$ 2.7 & B & 1,2,3\\
LHS\,511/Gl\,831         &	  45.1$\pm$ 1.8 & -73.3$\pm$ 1.6 & -43.6$\pm$ 1.3 &  -0.7$\pm$ 1.7 &  85.3$\pm$ 2.7 &   & 4\\
LHS\,515                 &	 267.9$\pm$55.0 &-107.2$\pm$37.2 &-126.1$\pm$29.8 &-234.9$\pm$26.7 & 287.4$\pm$55.5 &   & 4\\
LHS\,523                 &	  55.9$\pm$ 3.4 &  22.8$\pm$ 2.2 & -57.5$\pm$ 3.1 &  -6.6$\pm$ 3.5 &  62.2$\pm$ 5.2 &   & 4\\
LHS\,68/Gl\,866AB        &	  52.6$\pm$ 0.6 & -73.9$\pm$ 1.0 &  -8.1$\pm$ 1.5 &  25.8$\pm$ 2.2 &  78.7$\pm$ 2.9 &   & 4\\
  \end{tabular}
\raggedright
\noindent
\vskip 10pt
NOTES:\\
$a$ -- Note that the uncertainty ellipse associated with the {\em measured} space velocities does not have axes
orthogonal to the (U,V,W) co-ordinate system. This means that the uncertainties in the U,V,W velocities are not
independent, and so they do not add in quadrature to give the uncertainty in the total space velocity.\\
$b$ -- Magnitude of the total space velocity, corrected for the basic solar motion as described in the text.\\
$c$ -- {\bf Sample A} is the set of photometrically selected objects with I--K$>$3.5 (ie M$_{Bol}$$>$12.8). It also
includes the binary companion TVLM\,513-42404B, which is assumed to have the same radial velocity, distance and
proper motion as TVLM\,513-42404A. {\bf Sample B} is the set of proper motion selected objects with I--K$>$3.5 (ie M$_{Bol}$$>$12.8).\\
ASTROMETRY REFERENCES:\\
(1) Tinney \etal\ 1995; (2) Tinney 1996; (3) Monet \etal\ 1992; (4)  van Altena, Lee \& Hoffleit 1995.
\label{uvw}
\end{table*}

\begin{table*}
\center
\caption{Mean Velocities and Dispersions for VLM Stars.}
\begin{tabular}{ccccccccccc}
          &\multicolumn{3}{c}{Weighted}
                                &&\multicolumn{3}{c}{Unweighted}
                                                       && \multicolumn{2}{c}{Early M dwarfs}  \\
              &  Sample A   &  Sample B   & SampleA+B &&  Sample A  &  Sample B     &Sample A+B &&   Hawley \etal$^a$\\
              &  (km/s)   &  (km/s)   & (km/s)              &&  (km/s)  &  (km/s)   & (km/s)  &&  (km/s)  \\
\\
   $<$U$>$    & -13$\pm$7    &  -34$\pm$14 & -17$\pm$8 && -18$\pm$7  &  -34$\pm$14   & -25$\pm$7 &&-18.6\\
   $<$V$>$    & -20$\pm$6    &  -42$\pm$11 & -30$\pm$6 &&  -27$\pm$6 &  -42$\pm$11   & -33$\pm$6 &&-32.1\\
   $<$W$>$    &  -8$\pm$4    &  -13$\pm$4  & -10$\pm$3 &&   -8$\pm$4 &  -13$\pm$4    & -10$\pm$3 &&-13.6\\
          &\\
${\sigma}_U$&    18         &    34       &    26      &&   25      &   45          &    35     && 38.2\\
${\sigma}_V$&    16         &    30       &    25      &&   22      &   34          &    28     && 25.7\\
${\sigma}_W$&    23         &    30       &    25      &&   25      &   34          &    27     && 20.5\\
\\
\end{tabular}
\raggedright
\noindent
\vskip 10pt
NOTES:\\
$a$ - The ``Complete'' sample results from Table 9 of HGR, but corrected for basic solar motion as for our
data (\S3). 
\label{meandisp}
\end{table*}

Our velocities for BRI\,0021-0214 show marginal evidence for variability.
Moreover, they are also significantly different from that obtained by Basri \& Marcy
in November 1993. Multiple observations on that run ruled out velocity variability on 
90 minute timescales at the 3\,km/s level (indicating the ``system'' is not
a close, tidally-locked binary). The two combined sets of data therefore
suggest $\sim 2-12$ month velocity variability at the 10\,km/s level. Further
observations are required to confirm or reject this hypothesis. It should also be noted
that both we (\S6) and Basri \& Marcy find BRI\,0021-0214 to be a rapid rotator
($v{\sin}i \approx 40$\,km/s. This means its cross-correlation peak is significantly
broadened, making it possible that the velocity variability observed is simply
due to the uncertainty in measuring the centroid of a broadened peak.

With the exception of these two objects, our velocities are consistent with
external observations and we conclude that they are 
accurate to the stated precisions. In particular, we are now confident that our
re-reduction of the Palomar radial velocity data provides
a robust velocity system, consistent with our ESO data and with
independent observations. {\em The data presented in Tables \ref{table_phot} and \ref{table_prop}, therefore,
supersede that presented in RTM, which should no longer be used.}

\section{Radial Velocities \& Space Velocities}

The current sample includes 37 stars; 20 fainter than M$_{Bol} = 12.8$ (I--K $>$ 3.5) and 17 brighter
(the two stars with no I--K colour data are clearly more luminous than M$_{Bol} = 12.8$, given their
a spectral types of M5.5V and sdM2).  A subset of these stars were
identified originally through their substantial proper motions, and
one might expect that subset to be biased toward higher velocities.
To take this into account, we define two sub-samples:
\begin{itemize}
\item[]{\bf Sample A} - is the set of photometrically selected objects with I--K$>$3.5 (M$_{Bol} > 12.8$),
                        containing 12 objects (Table \ref{table_phot});
\item[]{\bf Sample B} - is the set of proper motion selected objects with I--K$>$3.5 (M$_{Bol} > 12.8$),
                        containing 9 objects(Table \ref{table_prop}). 
\end{itemize}
Table \ref{uvw} lists the Galactic (U,V,W) space motions (U positive toward
the Galactic Centre), where we
correct for a Basic Solar Motion of (9,11,6)\,km/s (Mihalas \& Binney 1981).
In each case, the radial velocity and 
proper motion uncertainties were propagated into the (U,V,W) system. However, it should be noted
that since the uncertainties in U,V and W are not independent, they do not add in quadrature to give
the uncertainty in the total space velocity.

None of the ``confirmed'' radial velocity variables in Tables \ref{table_phot} and \ref{table_prop} (eg. Gl\,866AB) are included 
in Samples A or B. The possible variables BRI\,0021-0214 and TVLM\,832-10443 are included pending a final
determination of their status -- in any case their velocity variation is of the order of the quoted
uncertainty in their velocities.
The total number of objects in Samples A and B is only slightly larger than that
used in the equivalent samples of RTM (ie 21 versus 18), but the
precision is improved significantly and none of the current radial velocities
are derived from H$\alpha$ emission.
Altogether, the conclusions we derive from these new data can be expected to be
considerably more robust.

\begin{figure*}
\centerline{\psfig{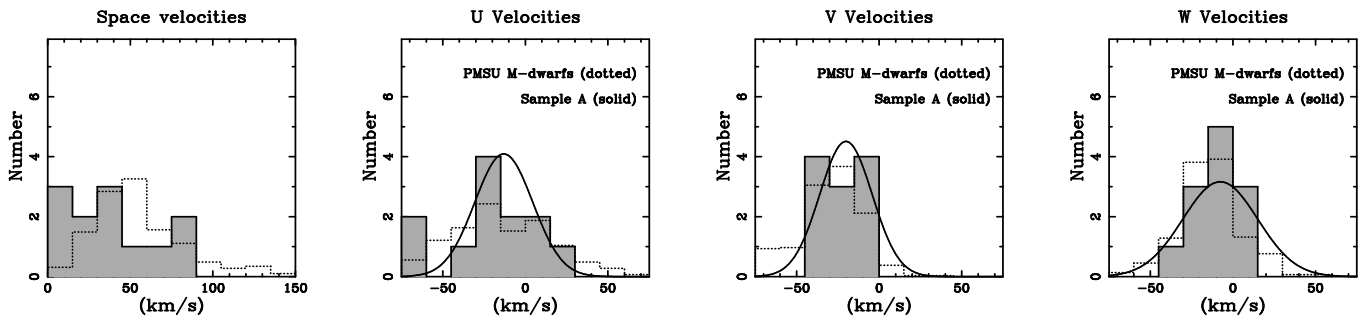}}
\caption{The V$_{total}$, U, V and W
velocity distributions for Sample A are shown as shaded. A solid Gaussian shows
the weighted mean and dispersion. The dotted histogram is the ``Complete''
sample of HGR, normalised to the same total counts as Sample A. 
All velocities have been corrected for the basic solar motion
as described in the text. }
\label{afig}
\end{figure*}

\begin{figure*}
\centerline{\psfig{file=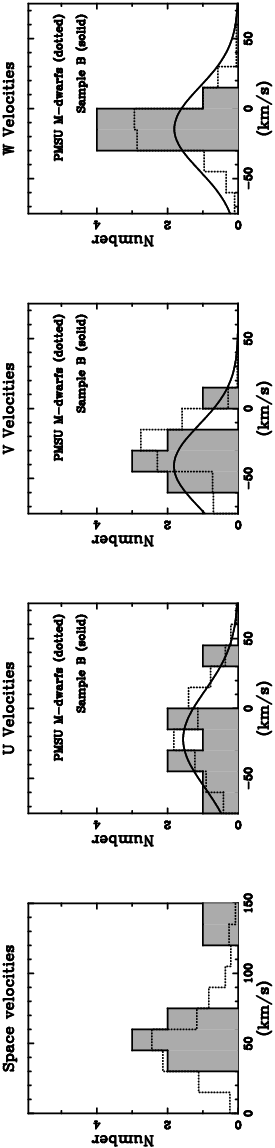,width=18cm,rotate=90}}
\caption{As for Fig. \ref{afig}, but for Sample B.}
\label{bfig}
\end{figure*}

\begin{figure*}
\centerline{\psfig{file=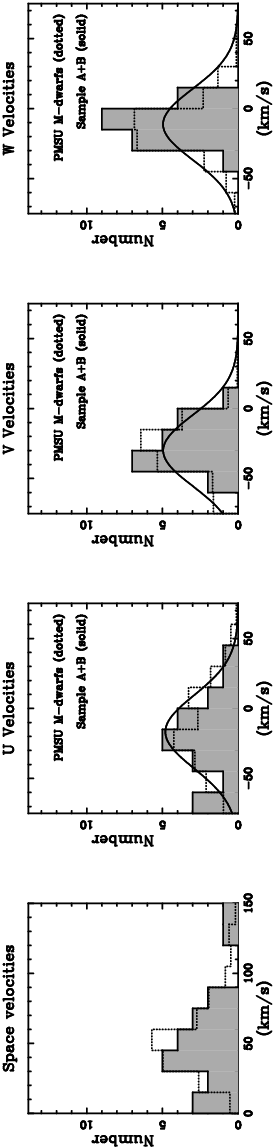,width=18cm,rotate=90}}
\caption{As for Fig. \ref{afig}, but for Sample A+B.}
\label{abfig}
\end{figure*}

Table \ref{meandisp} shows the means ($<$U$>$,$<$V$>$,$<$W$>$) and standard deviations 
(${\sigma}_U$,${\sigma}_V$,${\sigma}_W$) of the Galactic velocity distributions
for Samples A, B and A+B. Both weighted and unweighted estimates of these quantities are shown. 
The same quantities are shown for the 346 stars of the more luminous (M$_{V} \approx 8-13$, or 
M$_{Bol} \approx 7.4-10.4$) volume-complete sample of Hawley, Gizis \& Reid (1997 - hereafter HGR), 
with the same correction applied for basic solar motion.
Figures \ref{afig}-\ref{abfig} show the V$_{total}$, 
U, V, and W velocity
distributions for the same three samples, together with the corresponding distributions from HGR.


Inspection of Table \ref{meandisp} shows that whether weighting is used or not makes no
significant difference to the observed means -- which  are in all cases consistent 
with those observed for the more massive stars of HGR. 
The value of the velocity dispersion depends slightly on
whether weighting is used, with the weighted values being somewhat 
smaller. Again, however, there is no indication in any case that the velocity
dispersion for VLM stars is smaller than that seen for early M-dwarfs. 

\begin{table}
\center
\caption{K-S Test Results -- the probabilities, for each velocity component,
that the hypotheses described in the text are true.}
\begin{tabular}{cccccc}
 &\multicolumn{4}{c}{Hypothesis} \\
 & (a)  & (b)  & (c) & (d)\\
Velocity\\
 U & 33\% &78\%& 67\% & 53\%\\
 V & 33\% &89\%& 77\% & 32\%\\
 W & 96\% &17\%& 12\% & 91\%\\
\\
\end{tabular}
\label{kstests}
\end{table}

Because such an ``eyeball'' comparison can be misleading, we have performed 
Kolmogorov-Smirnov tests (see eg. Press {\em et al.}\ 1986) to  
test the following hypotheses: 
\begin{itemize}
\item[(a)] that Samples A and B are drawn from the same statistical population; 
\item[(b)] that Sample A+B is drawn from the same statistical population as the HGR sample;
\item[(c)] that Sample A is drawn from the same statistical population as the HGR sample;
\item[(d)] that Sample B is drawn from the same statistical population as the HGR sample;
\end{itemize}
The results of these tests are shown in Table \ref{kstests} and
provide no evidence that samples A and B are drawn from different
parent populations. Figs \ref{afig}-\ref{abfig} suggest possible kinematic
differences  between the VLM-dwarf and HGR samples. However,
the K-S tests show that all of the samples are 
indistinguishable statistically.

\section{A Spectral Atlas at 0.43\AA\ Resolution}

\begin{figure*}
\centerline{\psfig{file=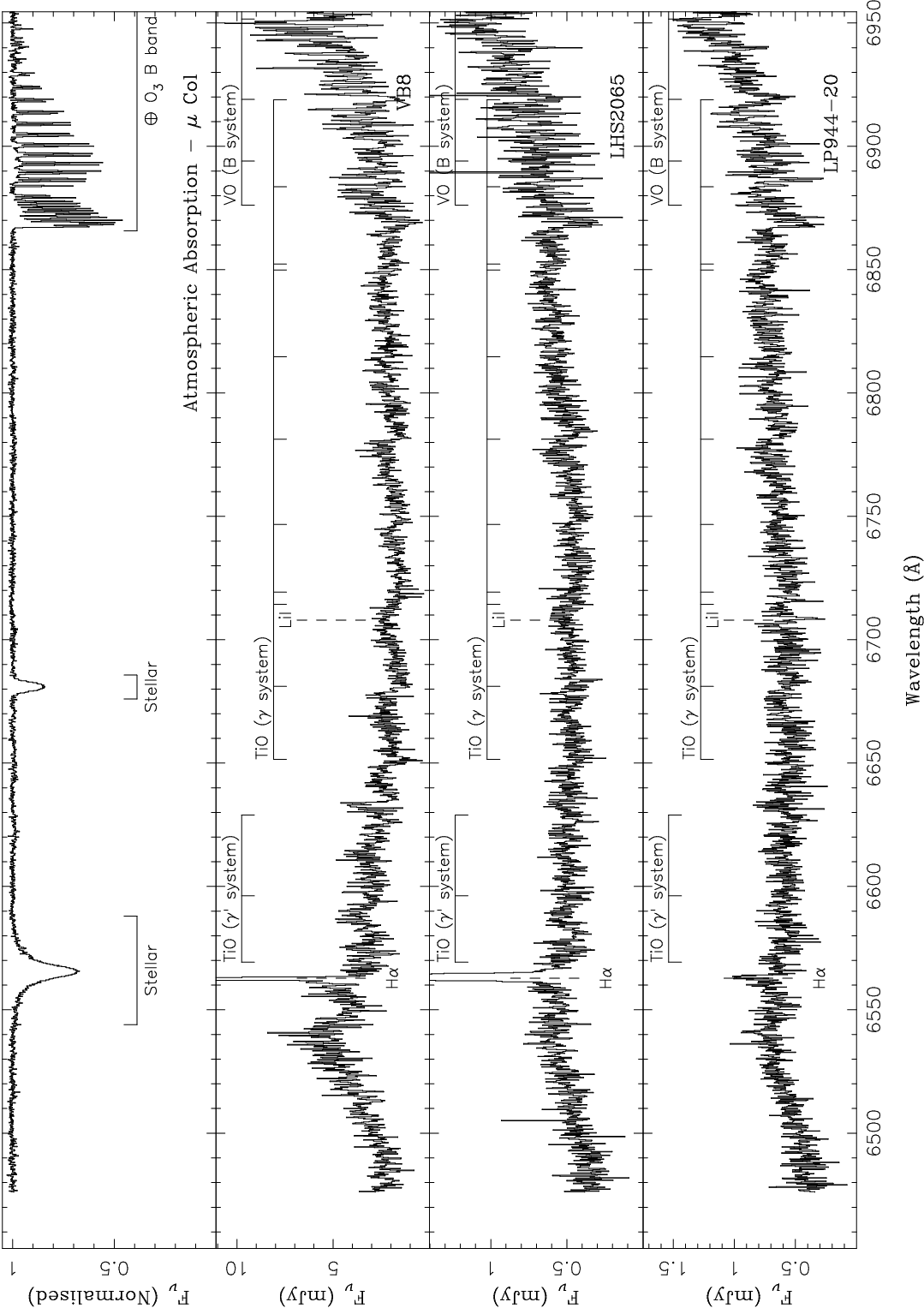,height=22.0cm}}
\caption{A spectral atlas for the objects VB\,8, LHS\,2065, and LP\,944-20
at 0.43\AA\ resolution. The normalised spectrum of the
standard star $\mu$\,Col shows the location of terrestrial
atmospheric features. The bottom of the scale in each panel is 0.0.}
\label{atlasstart}
\end{figure*}

High resolution echelle spectrographs efficient enough to observe faint targets
have been used now for several years to study the lowest mass stars (eg. Basri \& Marcy 1995;
Schweitzer \etal\ 1996). Despite this the available atlases
for these objects are all based on data taken at low resolution 
(eg. Kirkpatrick \etal\ 1991, 18\AA; Turnshek \etal\ 1985, 8-12\AA). We therefore
present in Figures \ref{atlasstart}-\ref{atlasend} an atlas showing CASPEC spectra of three objects;
the M7 dwarf VB\,8 (Kirkpatrick \etal\ 1991); the M9 dwarf LHS\,2065 
(Kirkpatrick \etal\ 1991) and the $>$M9V brown dwarf LP\,944-20 
(Kirkpatrick \etal\ 1997; Tinney 1998). 

These spectra have been flux calibrated using
observations of the standard star $\mu$\,Columbae (Turnshek \etal\ 1990). Because
our observations were not made with a spectrophotometric slit, we only obtain
a relative inter-order calibration. An absolute calibration has been obtained by 
scaling to the published I-band photometry of these objects (Leggett 1992) using a
zero-magnitude flux of 2550\,Jy (Reid \& Gilmore 1984). The overall flux calibration
is accurate to $\pm$10\% -- however the inter-order calibration is much better
and good to a few percent.
Photon counting errors were carried throughout the reduction. The signal-to-noise
ratio (SNR) as a function of wavelength for our spectra is summarised in Fig \ref{snr}.
Digital copies of the atlas spectra (including the SNR spectra) 
have been deposited with the NASA Astrophysics Data Centre.

\begin{figure*}
\centerline{\psfig{file=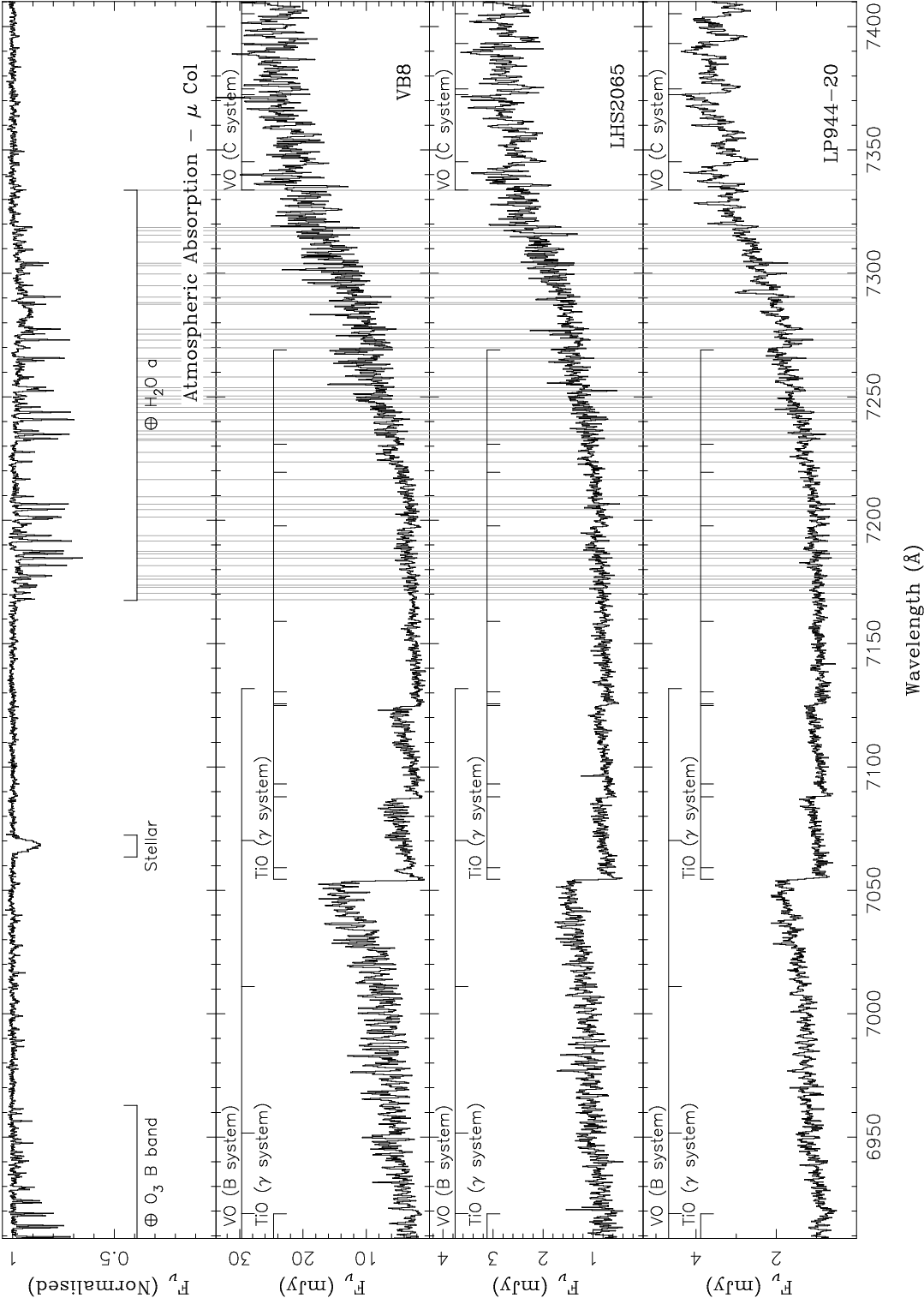,height=22.0cm}}
\caption{A spectral atlas for the objects VB\,8, LHS\,2065, and LP\,944-20
at 0.43\AA\ resolution (continued).}
\end{figure*}
\begin{figure*}
\centerline{\psfig{file=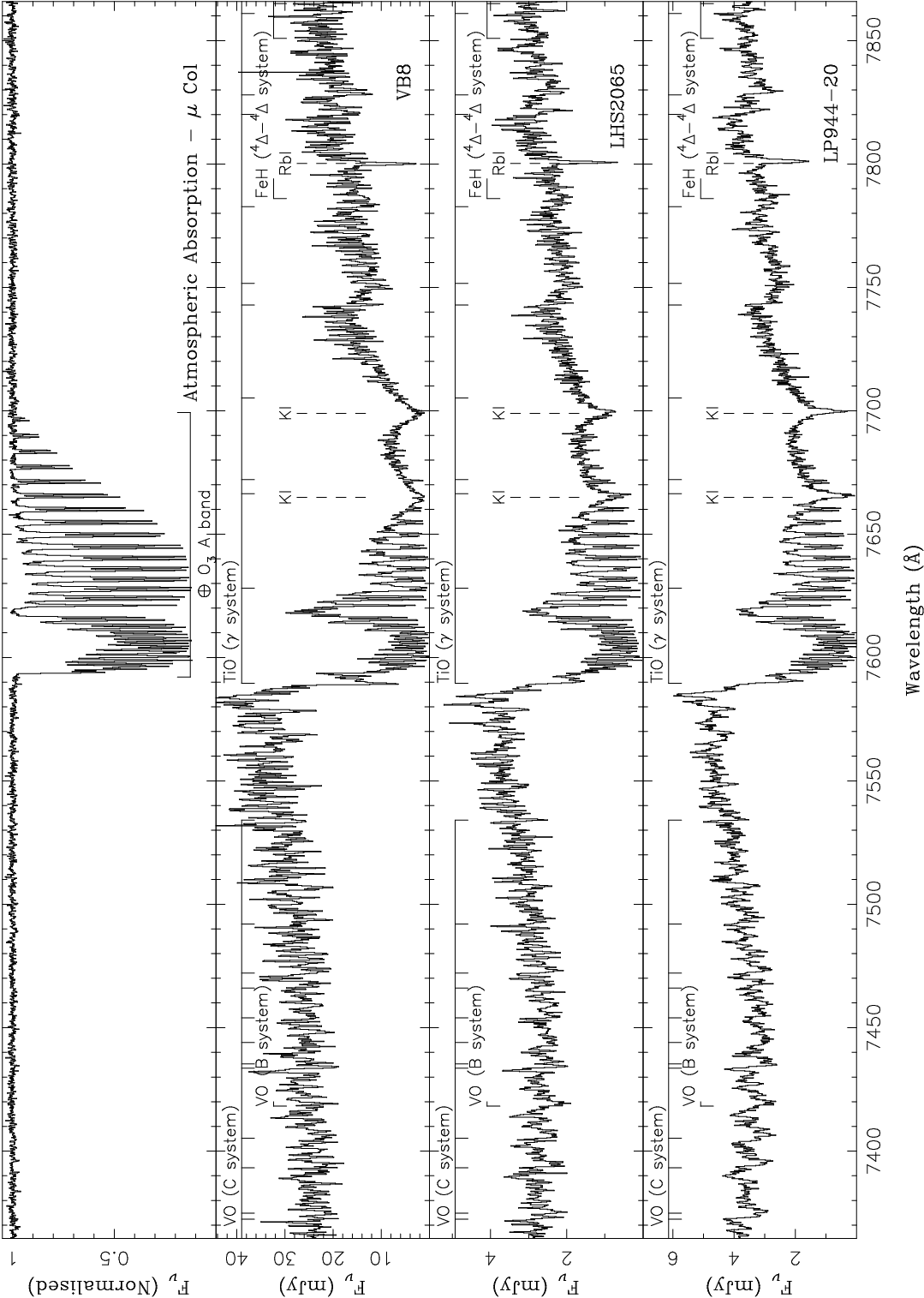,height=22.0cm}}
\caption{A spectral atlas for the objects VB\,8, LHS\,2065, and LP\,944-20
at 0.43\AA\ resolution (continued).}
\label{ki}
\end{figure*}
\begin{figure*}
\centerline{\psfig{file=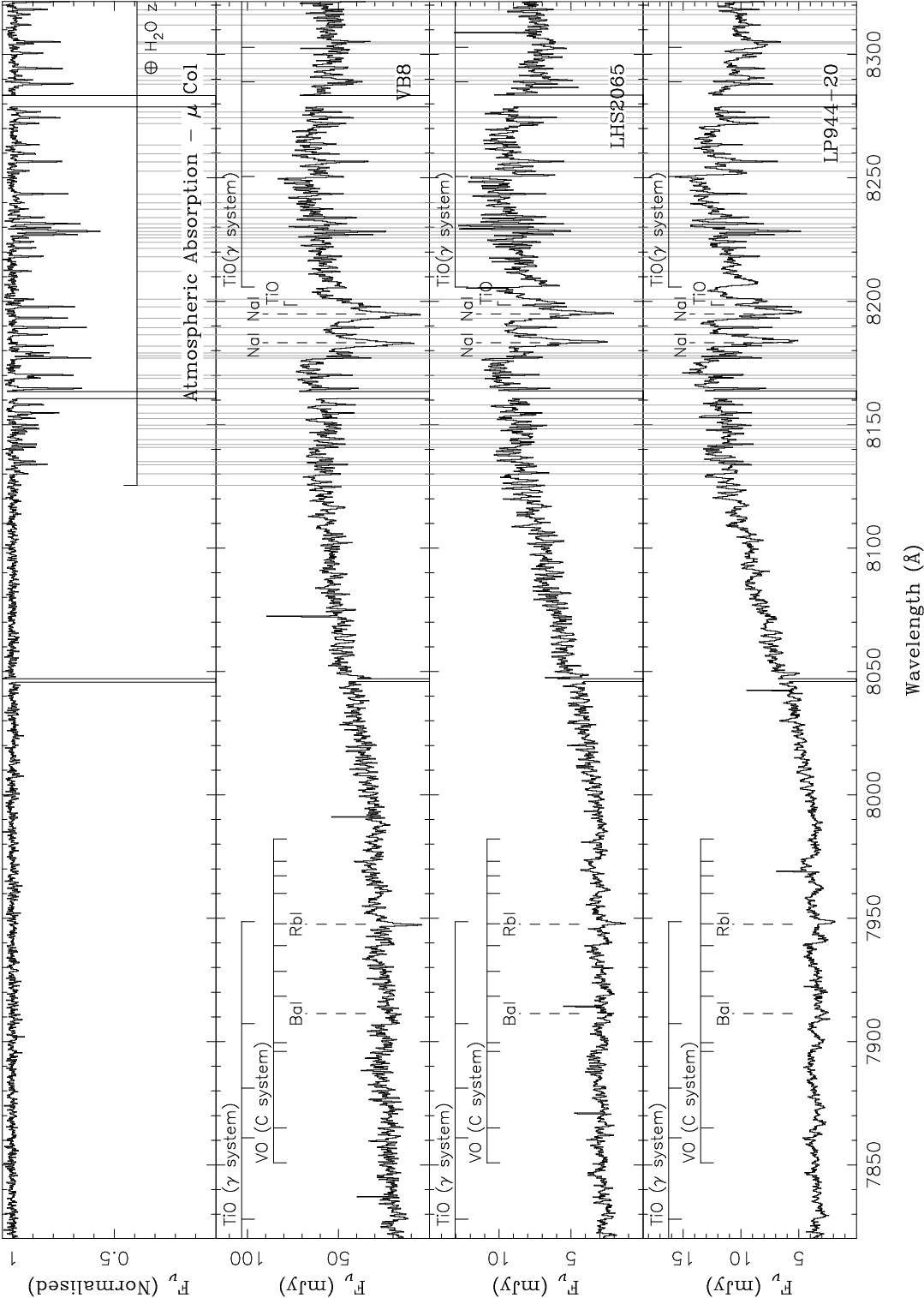,height=22.0cm}}
\caption{A spectral atlas for the objects VB\,8, LHS\,2065, and LP\,944-20
at 0.43\AA\ resolution (continued).}
\end{figure*}
\begin{figure*}
\centerline{\psfig{file=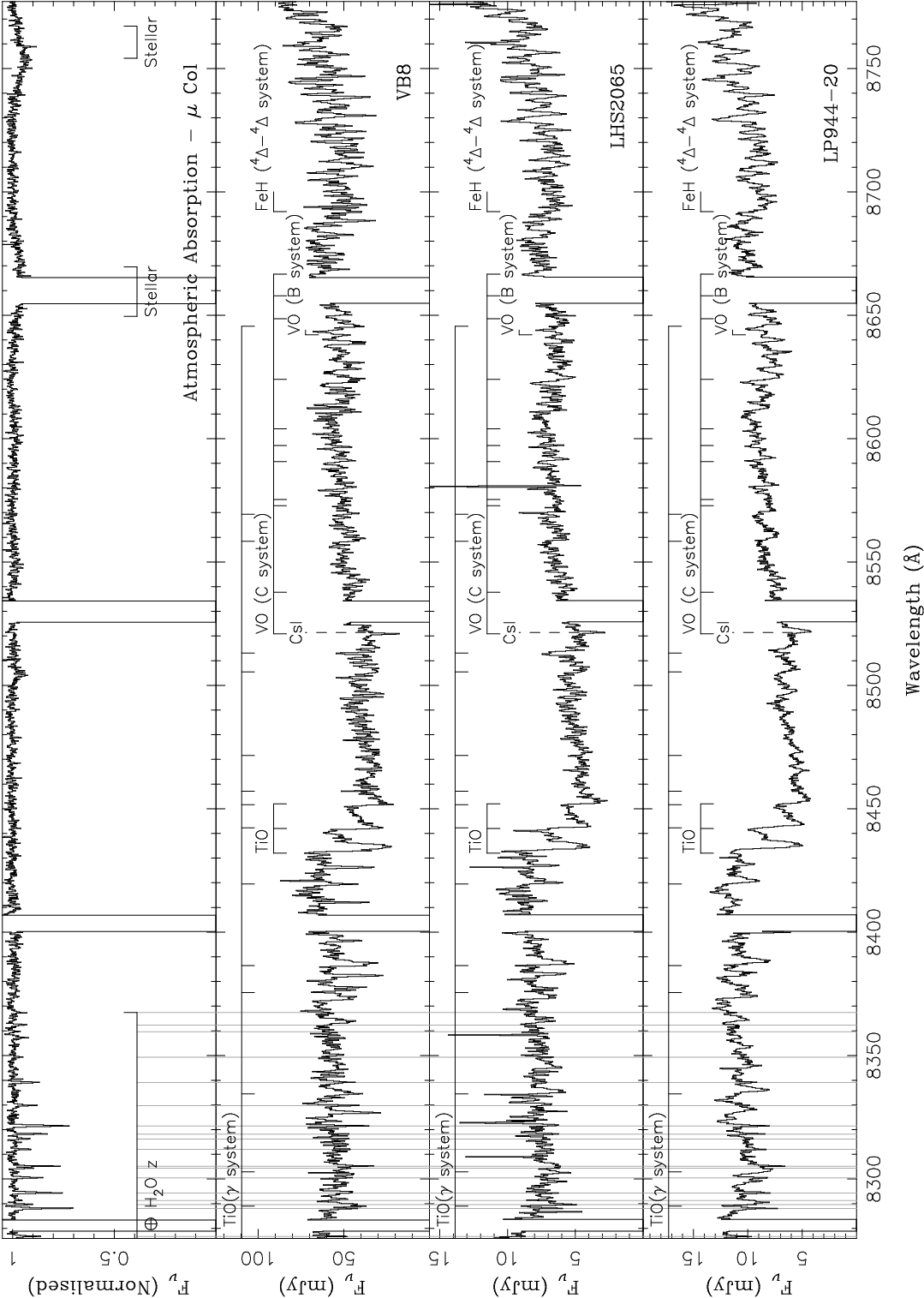,height=22.0cm}}
\caption{A spectral atlas for the objects VB\,8, LHS\,2065, and LP\,944-20
at 0.43\AA\ resolution (continued).}
\end{figure*}
\begin{figure*}
\centerline{\psfig{file=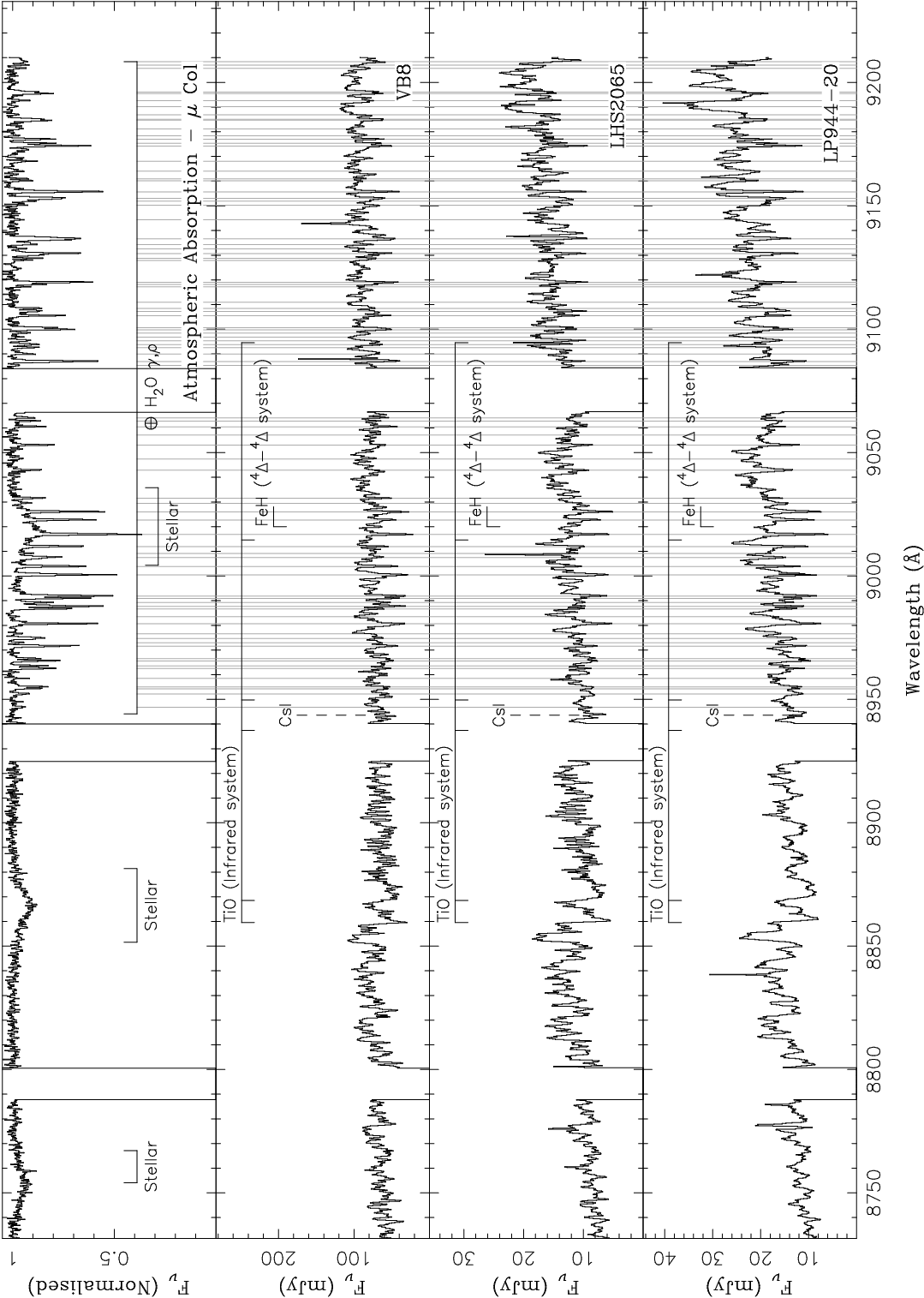,height=22.0cm}}
\caption{A spectral atlas for the objects VB\,8, LHS\,2065, and LP\,944-20
at 0.43\AA\ resolution (continued).}
\label{atlasend}
\end{figure*}

\subsection{Terrestrial Features}
The spectra shown have not been corrected for terrestrial absorption due to
O$_3$ and H$_2$O (cf. Allen 1976, \S58 and references therein). The locations
of terrestrial absorption in these spectra can be seen in the panel showing
the normalised flux from the standard star $\mu$\,Col. The
effects of O$_3$ are quite regular and straightforward to disentangle from
features in the stars themselves. This is not true for H$_2$O absorption which
is complex and irregular. We therefore show in the stellar panels the locations
of the strongest lines of H$_2$O absorption. Also shown are the locations of
the stellar absorptions present in $\mu$\,Col itself.

\subsection{Molecular Features}

The dominant molecular features in these spectra are due to TiO and VO. The
figure shows the locations of identified bandheads due to these
two species. Also shown are the spectroscopic ``systems'' to which these
bandheads have been assigned. Information on the actual state transitions
to which these systems correspond can be found in the references provided below --
in particular see Pearse \& Gaydon (1976).
It is important to emphasise that these identifications represent only the
bandheads -- more widely spread transitions 
cover almost all of the optical and infra-red spectra of these stars. Between 8700\AA\ and
8850\AA\ for example there are no major bandheads, but the spectra clearly
contain numerous molecular features, as can be seen in the
laboratory spectra of Gatterer \etal\ (1957).

\subsubsection{TiO} All the TiO bandheads are observed to be degraded to the red.
Where no reference is given consult Pearse \& Gaydon (1976), Solf (1978), Gatterer \etal\ (1957)
and Turnshek \etal\ (1985).
Indicated are: the $\gamma$ or ``Red'' system bandheads at 
6651.5, 6681.1, 6714.4, 6719.3, 
6746.7, 6781.3, 6814.7, 6849.9, 
6852.3, 6883.6, 6918.9, 7054.5, 7059.2, 7087.9, 7093.1, 7124.9, 7125.6, 7130.4, 7159.0, 
7197.7, 7219.4, 7230.8,
7269.0, 7589.6, 7628.1, 7666.4, 7672.1, 7705.2, 7742.8, 7751.6, 7782.7, 7820.1, 7828.0, 
7861.0, 7881.2,
7907.3, 7948.6, 8205.8, 8250.6, 8289.0, 8302.9, 8334.5, 8375.5, 8386.5, 8419.5, 8442.3, 8451.7, 8457.1, 8471.6, 8505.5, 8513.1, 8558.4, 8569.4, and 8645.5\AA; 
the $\gamma\prime$ or ``Orange-Red'' system bandheads at 
6569.3, 6596.2 and 6629.0\AA; 
an unnamed system with bandheads at 
8432, 8442 and 8452\AA\ (Solf 1978);
and an ``Infrared'' system with bandheads at 
8859.6, 8868.5, 8937.4, 8949.8, 9014.6 and 9094.5\AA\ (Philips 1950).
Also shown is a bandhead which fits into no system at 8198.5\AA\ (Solf 1978)

\subsubsection{VO} Unlike TiO which produces distinctive bandheads degraded to the red, VO in M-dwarfs
produces more diffuse absorption. Where no reference is given consult Pearse \& Gaydon (1976), 
Solf (1978), Gatterer \etal\ (1957) and Turnshek \etal\ (1985). Indicated are: the
B or ``Red'' system bandheads at 
6876.2, 6894.0, 6919.0, 6951.6, 7011.0,7070.2, 7131.7, 7418.2, and 8642.1\AA;  
the C or ``Infrared'' system bandheads at 
7333.8, 7345.3, 7372.4, 7374.8, 7393.2, 7405.2, 
7433.6, 7435.3, 7444.0, 7454.0, 7466.0, 7472.1, 7492.0, 7534.1, 7850.9, 7865.0, 7896.0, 
7899.6, 7918.4, 7928.5, 7938.9, 7947.7, 7960.1, 7967.2, 7973.1, 7982.1, 8520.9, 8537.7, 
8572.8, 8575.3, 8590.7, 8597.2, 8604.0, 8624.0, 8648.6, 8657.9 and 8666.6\AA. 

\subsubsection{FeH} FeH bandheads are seen degraded to the red. Indicated are the
$^4\Delta$-$^4\Delta$ system bandheads at 
7786, 8692 and 9020\AA\ (Philips \etal\ 1987).

\subsection{Atomic Features}

Atomic features shown are: H$\alpha$ 6562.8\AA; Li\,I 6708.0\AA; K\,I 7664.9\AA\ and 7699.0\AA;
Rb\,I 7800.2\AA\ and 7947.6\AA; Na\,I 8183.3\AA\ and 8194.8\AA; Cs\,I 8521.4\AA\ and 8943.6\AA;
Ba\,I 7911.3\AA.
With the exception of H$\alpha$ all these lines are resonance transitions. The weak lines
of Ca\,II, Mg\,I and Ti\,I seen in earlier M-dwarfs (Kirkpatrick \etal\ 1991) are
undetectable amongst the morass of molecular absorptions in these M7-M9$+$ dwarfs.
The Li\,I detection seen in LP\,944-20 in Fig. \ref{atlasstart} is discussed in Tinney (1998).

It is particularly interesting to note the pronounced width of the K\,I lines in VB\,10 compared with
the two cooler objects. Schweitzer \etal\ (1996) have modelled the atomic lines of VB10, and find that
this extreme strength of K\,I could be due to either high metallicity ([M$/$H]=+0.5) or high gravity
(log\,$g$=5.5). Schweitzer \etal\ did not explore the temperature dependence of K\,I line strength,
however the spectra shown in Fig. \ref{ki} would seem to imply that either K\,I weakens slightly 
with decreasing effective temperature, or that VB\,10 has an unusually high metallicity or gravity.


\begin{figure}
\centerline{\psfig{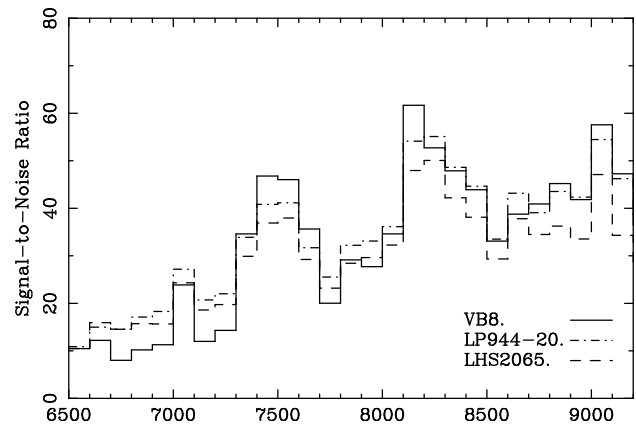}}
\caption{Signal-to-noise ratios as a function of wavelength for the spectra
shown in Figs \ref{atlasstart}-\ref{atlasend}.}
\label{snr}
\end{figure}

\section{Spectral Features}

\subsection{Atomic Lines}

In this section we deal only with the high resolution observations made at La Silla.
Na\,I and K\,I lines are prominent in almost all of these spectra, however measurement
of them is made difficult by the presence of atmospheric absorption, which
we did correct for this study.

Cs\,I and Rb\,I lines, however, lie in regions of the spectrum clear of
atmospheric absorption, and
become prominent for late-type dwarfs (Tinney 1997). The measured equivalent
widths of these lines are shown in Table \ref{csandrb}. Because no part 
of the spectrum 
is free of stellar molecular absorption, the assignment of a true
continuum is impossible. As a result whenever we refer to equivalent widths, 
they are always defined relative to the apparent 
continua available near the lines. 
The uncertainties presented in the table are based on photon-counting errors 
and do not include systematics due to pseudo-continuum placement. However,
our measurements should be self-consistent within this set of observations
and with other observations obtained at similar resolution.
We have also included results from Mart\'{\i}n \etal\ 1997 for two cool objects
discovered in the DENIS survey (Delfosse \etal\ 1997; Tinney, Delfosse \& Forveille 1997).

\begin{table}
  \caption{Cs\,I and Rb\,I Observations}
  \begin{tabular}{lllll}
                    &      &\multicolumn{3}{c}{Pseudo-Equivalent Widths (\AA)}  \\
Object              & I--K &Rb\,I 7800\AA &  Rb\,I 7948\AA &  Cs\,I 8521\AA    \\[5pt]
LP\,944-20          & 4.58 & 0.87$\pm$0.05 & 0.73$\pm$0.05 & 0.48$\pm$0.02\\
BRI\,1222-1222      & 4.31 &0.88$\pm$0.03 & 0.83$\pm$0.05 & 0.39$\pm$0.03\\
ESO\,207-61         & 4.18 &0.94$\pm$0.06 & 1.03$\pm$0.14 &    -$^a$  \\
LHS\,2065           & 4.46 &0.81$\pm$0.03 & 0.62$\pm$0.05 & 0.33$\pm$0.05\\
LHS\,2351           & 3.65 &0.76$\pm$0.05 & 0.63$\pm$0.05 & 0.30$\pm$0.03\\
LHS\,2397a          & 4.26 &0.87$\pm$0.02 & 0.96$\pm$0.05 & 0.50$\pm$0.10\\
LHS\,2875           & 2.05 &0.16$\pm$0.02 &    -       &    -      \\
LHS\,2876           & 3.59 &1.20$\pm$0.20 & 1.20$\pm$0.10 &    -      \\
LHS\,3003           & 3.66 &0.84$\pm$0.02 & 0.86$\pm$0.03 & 0.34$\pm$0.02\\
LHS\,427            & 2.30 &0.23$\pm$0.01 & 0.17$\pm$0.01 & $<$0.03   \\
LHS\,428            & 2.16 &0.19$\pm$0.01 &    -       & $<$0.03   \\
LHS\,429            & 3.42 &1.02$\pm$0.02 & 0.95$\pm$0.03 & 0.36$\pm$0.03\\
LHS\,474            & 3.97 &0.93$\pm$0.05 & 0.93$\pm$0.05 & 0.37$\pm$0.07\\[5pt]
DENIS-P\,J1058$^b$
                    & 5.1  &    -      &    -       & 2.1\\
DENIS-P\,J1228$^b$
                    & 5.5  &    -      &    -       & 4.3\\
  \end{tabular}
\raggedright
\noindent
\vskip 10pt
NOTES:\\
$a$ -- Cs\,I 8521\AA\ was detected in ESO\,207-61, but could not be measured because its apparent radial
       velocity moved the Cs\,I\,8521\AA\ line onto the end of an echelle order.\\
$b$ -- I--K from Delfosse \etal\ 1997. Equivalent widths for Cs\,I from Mart\'{\i}n \etal\ 1997. 
M$_{bol}$ estimates from Delfosse \etal\ 1997.
\label{csandrb}
\end{table}

\begin{figure}
\centerline{\psfig{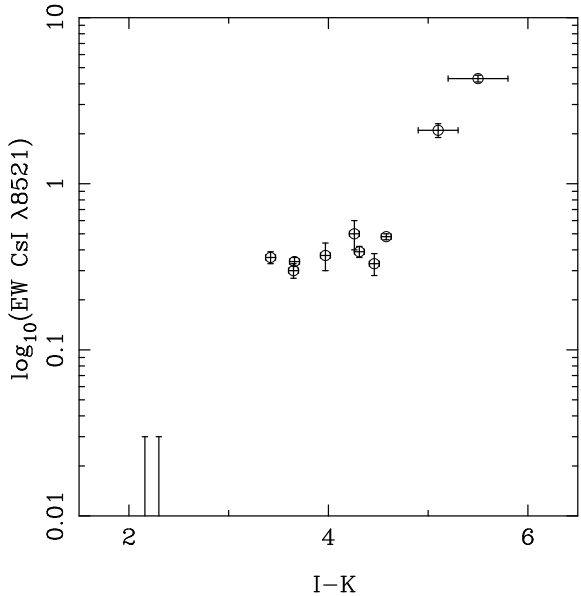}}
\caption{Strength of Cs\,I lines as a function of I--K colour for the objects in
Table \ref{csandrb}.}
\label{csfig}
\end{figure}

Cs I line strengths are plotted in Figure \ref{csfig}. 
The observations describe a plateau at EW=0.2-0.3\AA\ throughout
the late M-dwarf range, with considerably weaker absorption for the early M-dwarfs (as
represented by the two upper limits), and considerably stronger absorption for
objects below $\sim$2000K. This diagram highlights the possible utility of the Cs\,I\,8521\AA\
line as a low-resolution spectroscopic indicator of effective temperature -- at least
for the purpose of discriminating extremely cool objects like DENIS-P\,J1228 and DENIS-P\,J1058
(for which it has been proposed that an entirely new spectral type (``L'') is required; 
Kirkpatrick 1998) from late-type M dwarfs.

\subsection{H$\alpha$ Emission}

\begin{table}
  \caption{CASPEC H$\alpha$ Emission Measurements}
  \begin{tabular}{lccc}
Object              &     UT               & H$\alpha$ EW$^a$ & $[\frac{{\rm L}_{\rm H\alpha}}{{\rm L}_{\rm bol}}]^b$\\
                    &                      & (\AA)   \\[5pt]
BRI\,0021-0214      & 1993/6/28 07:55:22 &  $<$2        & $< -$5.1\\
LP\,944-20
                    & 1993/6/28 09:53:58 &  $<$2        & $< -$5.1\\
                    & 1994/2/09 01:15:52 &  0.6$\pm$0.2:& $-$5.6\\
BRI\,1222-1222      & 1994/2/08 09:00:17 &  4.7$\pm$0.5 & $-$4.7\\
ESO\,207-61         & 1994/2/08 02:17:06 &  2.8$\pm$0.5 & $-$4.9\\
                    & 1994/2/09 03:04:46 &  2.6$\pm$0.8:& $-$5.0\\
LHS\,2065           & 1994/2/08 04:22:29 &  24.9$\pm$0.3& $-$4.0\\
LHS\,2351           & 1994/2/09 05:44:56 &  6.1$\pm$0.1 & $-$4.3\\
LHS\,2397a          & 1994/2/08 06:13:18 &  47.3$\pm$0.8& $-$3.7\\
                    & 1994/2/09 06:59:29 &  34.6$\pm$1.0& $-$3.9\\
LHS\,2875           & 1994/2/09 07:59:21 &  0.20$\pm$0.02a\\
LHS\,2876           & 1994/2/09 08:40:30 &  1.6$\pm$0.5 & $-$4.9\\
LHS\,3003           & 1994/2/09 07:35:41 &  4.1$\pm$0.3 & $-$4.6\\
LHS\,427            & 1994/2/08 09:37:38 &  0.24$\pm$0.01a \\
LHS\,428            & 1993/6/28 03:04:22 &  1.8$\pm$0.4 & $-$3.5\\
                    & 1994/2/08 09:42:51 &  1.6$\pm$0.1 & $-$3.6\\
LHS\,429            & 1993/6/28 02:17:11 & 20.4$\pm$1.3 & $-$3.7\\
                    & 1994/2/08 09:25:37 & 3.7$\pm$0.2  & $-$4.5\\
                    & 1994/2/09 09:31:43 & 4.1$\pm$0.1  & $-$4.4\\
LHS\,474            & 1993/6/28 03:32:50 &  4.4$\pm$0.7 & $-$4.7\\
  \end{tabular}
\raggedright
\noindent
\vskip 10pt
NOTES:\\
$a$ -- ``$<$'' indicates a 3-$\sigma$ upper limit. ``:'' indicates
       a marginal detection. ``a'' indicates the line was detected in absorption. 
       Other results are actual measurements.\\
$b$ -- Upper limits are indicated by $<$. Measurements have an uncertainty of $\pm$0.2
       largely due to the calibration from equivalent width to [L$_{\rm H\alpha}$$/$L$_{bol}$].
\label{hatab}
\end{table}

\begin{figure}
\centerline{\psfig{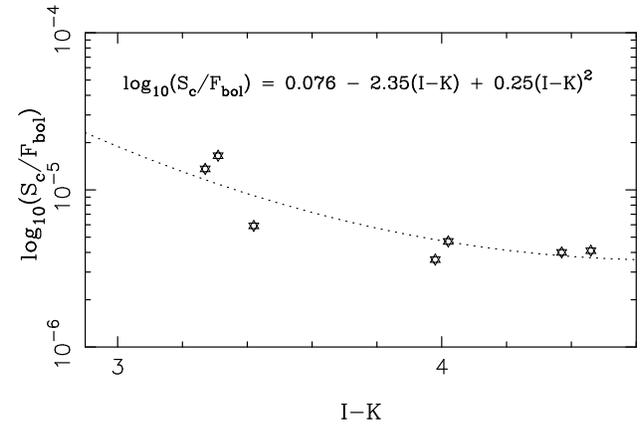}}
\caption{H$\alpha$ ``bolometric correction'' as a function of I--K colour for
VLM stars from Tinney, Mould \& Reid 1993. Also shown is the calibration
relation adopted to derive the [L$_{\rm H\alpha}$$/$L$_{bol}$] values derived
in Table \ref{hatab}}
\label{hafig}
\end{figure}

H$\alpha$ emission is seen in low mass stars when magnetic fields heating
the outer atmosphere producing temperatures of up to $\sim$10$^6$\,K in
the outer corona. Closer to the photosphere the chromosphere shows lower temperatures 
of $\simlt$10$^4$\,K, where resonance lines of Ca\,II and Mg\,II, and the 
hydrogen Balmer lines are produced (HGR).
H$\alpha$ emission equivalent widths for our Palomar observations were reported in RTM. 
Table \ref{hatab} lists similar measurements for our ESO program objects, 
including only those observations yielding either detections or
significant upper limits. Because of poor seeing during our 1993 
July observations, these come mostly from 1994 February. 

Equivalent width (EW) is a measure of line flux (F$_{\rm H\alpha}$), 
measured relative to the local continuum flux density (S$_c$).
$$ {\rm EW} \approx \frac{{\rm F}_{\rm H\alpha}}{{\rm S}_c} $$
A large equivalent width does not necessarily imply strong
chromospheric activity.  A better indicator is the ratio of the 
luminosity in the H$\alpha$
line to the total bolometric luminosity of the star, L$_{\rm H\alpha}$$/$L$_{bol}$,
or its common logarithm([L$_{\rm H\alpha}$$/$L$_{bol}$]).
Since our observations were not made with a spectrophotometric wide
slit, line fluxes could not be measured directly. However, the EW 
can be converted to L$_{\rm H\alpha}$/L$_{\rm bol}$
if the ratio of the local continuum flux density S$_c$ to the
bolometric flux (F$_{\rm bol}$) is known. Since,

$$ \frac{{\rm L}_{\rm H\alpha}}{{\rm L}_{\rm bol}} = \frac{{\rm F}_{\rm H\alpha}}{{\rm F}_{\rm bol}} 
= \frac{{\rm EW} . {\rm S}_c}{{\rm F}_{\rm bol}} $$

The ratio S$_c$/F$_{\rm bol}$ is 
just a bolometric correction. We can therefore derive a functional
form for it, dependent on effective temperature or bolometric luminosity.
We have calculated this quantity for 
the sample of objects with calibrated optical spectra
and bolometric fluxes given in Tinney, Mould \& Reid (1993), and
the results are shown in Fig \ref{hafig}. The mean calibration 
was then used to derive the [L$_{\rm H\alpha}$$/$L$_{bol}$]
estimates shown in Table \ref{hatab}, with an estimated uncertainty
of $\approx \pm$0.2. Full discussion of these results is
deferred to \S7.

\section{Rotation}

Estimates of the rotational broadening of our CASPEC target stars have 
been obtained by cross-correlating the target spectra against a template 
({\sl cf.} Tonry \& Davis, 1979). The dispersion\footnote{In this 
section we use the term ``dispersion''
to describe the quantity $\sigma$ in the usual Gaussian parameterisation $f(x) \propto e^{-(x/\sigma)^2}$,
which is related to the full width at half maximum (FWHM) by $\sigma \approx {\rm FWHM}/2.354$.} 
of the cross correlation peak ($\mu$) is then the
quadratic sum of the dispersion of the instrumental broadening in each spectrum ($\tau$ -- assumed
the same in both), any intrinsic
broadening in the template spectrum ($\sigma_0$) and the rotational broadening 
of the target spectrum ($\sigma$). Uncertainties in the measurement of $\mu$ 
render it
impossible to measure values of $\sigma < \sqrt{2\tau^2 + \sigma_0^2}$. 
In the case of our CASPEC observations, $\tau = 6.8$\,km/s, setting a
detection limit for 
rotational broadening ($v{\sin}i$) of 9.6\,km/s. The Palomar data have
a FWHM resolution of 75\,km/s and set no useful limits on stellar rotation.

Following Tonry \& Davis we estimate the width $w$ of the cross-correlation peak
$\mu$ by fitting a parabola. However, unlike Tonry \& Davis we have not sought to
calibrate the relationship between this parabola width $w$ and the Gaussian dispersion
$\mu$. Rather we derive an empirical calibration between $w$ and $v{\sin}i$ by
cross-correlating  ``spun-up'' spectra of VB\,8 against an
independent observation of the same star. The rotationally-broadened spectra
are constructed using the profiles
given by Gray (1992, equation 17.12, $\epsilon=0.5$), computed for a series
of velocities between 0 and 50\,km/s.
The resulting $w$-to-$v{\sin}i$ calibration was calculated independently for each order.

Our calibration assumes that VB\,8, the template for our VLM sample, has
negligible rotation and there are no previous observational constraints of
this parameter. As a test, we have repeated our $w$-$v{\sin}i$ calibration
procedure using Gl\,643, which has a measured rotational upper limit
of $v{\sin}i < 2.7$\,km/s (Delfosse et al, 1998), cross-correlating
against the VB\,8 template. The resulting relation is essentially identical
to that derived using VB\,8 alone, implying that the latter star has a
rotation substantially lower than our instrumental resolution of 
$\tau = 6.8$\,km/s.


\begin{table}
  \caption{CASPEC Rotational Measurements}
  \begin{tabular}{lcc}
Object              &     UT               & $v {\sin} i^a$     \\ 
                    &                      & (km/s)   \\[5pt]
BRI\,0021-0214      & 28 Jul 1993 07:55:22 &  42.5$\pm$8.4\\
                    & 29 Jul 1993 08:36:41 &  40.1$\pm$8.6\\
LP\,944-20/BRI\,0337-3535$^b$
                    & 28 Jul 1993 09:53:58 &  28.3$\pm$1.9\\
                    & 29 Jul 1993 10:17:28 &  28.4$\pm$3.9\\
                    & 09 Feb 1994 01:15:52 &  28.1$\pm$1.8\\
BRI\,1222-1222      & 08 Feb 1994 09:00:17 &  $<$17\\
BR\,2339-0447$^d$   & 28 Jul 1993 06:22:05 &  $<$22\\
ESO\,207-61         & 08 Feb 1994 02:17:06 &  $<$18\\
                    & 09 Feb 1994 03:04:46 &  $<$19\\
LHS\,234/Gl\,283B   & 09 Feb 1994 04:25:04 &  $<$14\\
LHS\,2065           & 08 Feb 1994 04:22:29 &  $<$18\\
LHS\,2351           & 09 Feb 1994 05:44:56 &  $<$16\\
LHS\,2397a          & 08 Feb 1994 06:13:18 &  23.1$\pm$3.2$^b$\\
                    & 09 Feb 1994 06:59:29 &  23.6$\pm$1.8$^b$\\
LHS\,2875           & 29 Jul 1993 23:42:57 &  $<$22\\
                    & 09 Feb 1994 07:59:21 &  $<$17\\
LHS\,2876           & 09 Feb 1994 08:40:30 &  $<$18\\
LHS\,3003           & 09 Feb 1994 07:35:41 &  $<$13\\
LHS\,427/Gl\,643    & 28 Jul 1993 02:50:07 &  $<$19\\
                    & 08 Feb 1994 09:37:38 &  $<$18\\
LHS\,428/Gl\,644AB  & 28 Jul 1993 03:04:22 &  30.5$\pm$3.6$^c$\\
                    & 08 Feb 1994 09:42:51 &  $<$19\\
LHS\,474/Gl\,752B/VB\,10
                    & 28 Jul 1993 03:32:50 &  $<$19\\
                    & 29 Jul 1993 04:25:38 &  $<$19\\
LHS\,511/Gl\,831    & 28 Jul 1993 05:47:10 &  $<$19\\
  \end{tabular}
\raggedright
\noindent
\vskip 10pt
NOTES:\\
$a$ -- $<$ indicates a 3-$\sigma$ upper limit calculated as described in the text. Other results are
actual measurements.\\
$b$ -- The velocities shown are only marginally above the 3-$\sigma$ detection threshold for these observations.\\
$c$ -- Evidence was seen in the cross-correlation function for a double peak. This system is multiple so the
width measured is probably due to orbital motion, not rotation.\\
$d$ -- An M7-8III giant (Kirkpatrick, Henry \& Irwin 1997).\\
\label{rottab}
\end{table}

The rotational velocity for each target star was then estimated as follows: 
each order was cross-correlated with the appropriate template star order; the width $w$
of the highest peak in the cross-correlation was measured, if its confidence level exceeded
90\% (Tonry \& Davis, equation 33); the uncertainty in the measurement of this width ${\Delta}w$ was
also estimated (Tonry \& Davis, equation 25); if the measured width was less than $w_0 + 3{\Delta}w$, 
then this measurement is treated as an upper limit at the $v{\sin}i$ corresponding to $w = w_0 + 3{\Delta}w$,
otherwise the measured width was treated as a velocity measurement at the appropriate $v{\sin}i$;
and lastly, the resulting rotation measures were averaged over the available orders.

The results of this process are shown in Table \ref{rottab}. Typical uncertainties in  ${\Delta}w$ 
range from 0.2\,km/s
for bright stars to 2\,km/s for faint ones. This results in 3-$\sigma$ rotation detection
thresholds ranging from 15 to 22\,km/s. In other words, given the resolution 
of our observations,
we are only sensitive to very fast rotators. Significant rotation was detected in only two
objects: BRI\,0021-0214 and LP\,944-20. LHS\,2397a seems to show possible evidence for rotation
right at the detection limit for this star of $\approx 23$\,km/s. Gl\,644AB showed some evidence
for broadening in 1993 July, but none in 1994 February. Given the multiple nature of this system,
and evidence seen for a double peak in the cross correlation function, 
it seems reasonable to assume that the broadening observed in 1993 is more likely due to orbital
motions than stellar rotation. The effect of rapid rotation can be seen in the spectra
for LP\,944-20 plotted in Figures \ref{atlasstart}-\ref{atlasend}, where both atomic and molecular features are noticeably broader
than those observed in LHS\,2065 or VB\,10.
Of the two stars in which we detect rotation, only BRI\,0021-0214 has been previously
measured. The value of $41\pm8$\,km/s we measure for
BRI\,0021-0214 compares favourably with the 40$\pm7$\,km/s measured by Basri \& Marcy (1995).

\section{Rotation \& Activity}

The standard picture of the chromospheric activity (as revealed by H$\alpha$ emission) is
that it is a product of stellar magnetic fields, via the effects of stellar rotation
and convection combining to form a dynamo. In this picture, the activity is 
related directly to stellar rotation. Such a model is now well established for F,G and K-type
stars (Hartmann \& Noyes 1987). Stellar rotation rate is also a function
of age -- the rotation of a star slows as it loses angular momentum and ``spins down''.
The spin down rate is generally held to be a function of mass, 
with lower mass stars spinning down more
slowly (eg. Stauffer \etal\ 1994).

As a result a general relationship between chromospheric activity
and age can be deduced, at least for F, G and K dwarfs, though this relationship
is complicated by the fact that stars are
born with an initial spread in angular momentum, producing considerable scatter
about the overall relationship between chromospheric activity and age.
This is evidenced by the large spread seen in the rotational velocities of 
stars in young and intermediate-age clusters: for example, 
G dwarfs in the Pleiades ($\approx$100\,Myr) 
have $v{\sin}i$ $<$6 - 50\,km/s, while M dwarfs in the Hyades 
($\approx$600\,Myr)
show $v{\sin}i$ $<$10 - 20\,km/s (Stauffer \etal\ 1994).

The situation for low mass M dwarfs, however, is less clear. The appearance
of H$\alpha$ emission increases with spectral type from M0 ($\approx$ 10\%) to M6 ($\approx$ 50\%). 
It has been suggested that this is because H$\alpha$  becomes easier to detect
as the continuum flux at 6562\AA\ decreases.
However, the ratio of the H$\alpha$ luminosity to the bolometric luminosity 
(L$_{\rm H\alpha}$/$L_{\rm bol}$) spans an approximately constant range of
activity at [L$_{\rm H\alpha}$/L$_{\rm bol}$] = $-3.5$ to $-4.2$ (HGR),
rather than declining with decreasing luminosity, as one would expect
if a selection effect were responsible for the higher dMe fraction.

The straightforward interpretation of this result is that the
proportion of rapidly rotating M-dwarfs increases with spectral type.
Delfosse \etal\ (1998) have studied the rotation
of a sample of M-dwarfs, and find the expected correlation with rotation 
-- all stars with $v{\sin}i > 15$\,km/s show chromospheric activity at
the [L$_{\rm H\alpha}$/$L_{\rm bol}$] = $-3.5$ to $-4$ level, while stars with
no significant rotation detected show 
log$_{10}$[L$_{\rm H\alpha}$/L$_{\rm bol}$] = $-$4 to $-$4.5. Thus, this
result suggests that the higher fraction of late-type dMe dwarfs may
reflect longer spin-down times at very low masses.

However, the fast rotators detected in our CASPEC study throw a spanner
into this neat picture. Basri \& Marcy (1995) first pointed out that
BRI\,0021-0214 is extremely late in spectral type ($>$M9.5V, Kirkpatrick  \etal\ 1997), and a fast rotator,
{\em but} shows no H$\alpha$ in emission. Since then Tinney \etal\ (1997)
have found BRI\,0021-0214 to emit H$\alpha$ at very weak levels
(1.30$\pm0.05$\AA\ EW). This corresponds to only
 [L$_{\rm H\alpha}$/L$_{\rm bol}$] = $-5.3$ (cf. \S5.2) making this
to all intents and purposes a chromospherically inactive star, and directly contradicting
the generally held activity-rotation connection.

LP\,944-20, with [L$_{\rm H\alpha}$/L$_{\rm bol}$] $\approx -5.6$ and $v{\sin}i = 28$\,km/s,
represents the detection of a second member of
this class of ``inactive, rapidly rotating'' objects. Moreover, in this case we can add the further
physical parameters that the inactive, rapid rotator is a brown dwarf of mass
0.06\msol, and age 475-650\,Myr (Tinney 1998). The recently discovered brown dwarf
DENIS-P\,J1228.2-1547 ($v{\sin}i$=20\,km/s, [L$_{\rm H\alpha}$/L$_{\rm bol}$]$\approx$-5.3), 
and the brown dwarf candidate DENIS-P\,1058.7-1548 (23\,km/s, $\approx$-5.4) also
fall into this group (Mart\'{\i}n \etal\ 1997; Tinney \etal\ 1997). 
These objects may represent an entire class of
objects at or below the hydrogen burning main sequence which show substantially reduced
levels of chromospheric activity, despite the presence of levels of stellar rotations
which would produce significant activity in an M0-6 dwarf. 

Is this violation of the rotation-activity connection purely a function of
mass? Observations of nearby clusters suggest that this may be the case.
Activity (as measured by [L$_{\rm H\alpha}$/L$_{\rm bol}$]) amongst Pleiades
M-dwarfs rises monotonically with decreasing luminosity (and mass) until 
M$_{bol} \approx 10$, at which point the relative flux in H$\alpha$ drops
sharply. Figure 6(b), for example, of Reid, Hawley \& Mateo (1995) shows that for
luminosities below M$_{bol} \approx 10$, [L$_{\rm H\alpha}$/L$_{\rm bol}$]
falls from $\approx -3.5$ to $\approx -4.5$. Similar results have been derived by
Stauffer \etal\ (1994), Hodgkin, Jameson \& Steele (1995) and Oppenheimer \etal\
(1997), while Reid \& Hawley (1998) have detected comparable behaviour
amongst Hyades stars. 

Spectroscopy of Pleiades brown dwarfs shows that similar mass objects to LP944-20
and DENIS-P\,J1228
(Rebolo, Zapatero Osorio, Mart\'{\i}n 1995; Basri, Marcy \& Graham 1996; 
Zapatero Osorio \etal\ 1997) have H$\alpha$ lines with equivalent widths of $\sim 5-10$\AA, 
which corresponds to [L$_{\rm H\alpha}$/L$_{\rm bol}$]  $\approx$ $-4.5$ to $-4.8$ (\S 5.2)
at age $\sim 70-120$\,Myr.

We therefore conclude that (1) we see evidence in the Pleiades for
a trend towards lower chromospheric activity at masses below $\approx 0.1$\,\msol,
and (2) that we see no evidence for a significant change in this trend as the
brown dwarf limit is crossed.
Moreover, the Pleiades VLM stars studied by Oppenheimer \etal\ (1997) are all rapid rotators,
with $v{\sin}i$ in the range 37--65\,km/s. This implies that Pleiades age objects 
just above the H-burning limit already violate the rotation-activity connection.
The ``inactive rapid-rotators'' detected in the field continue this
violation below the H-burning limit. Age does not seem to play a {\em determining}
role, since both $\sim 70-100$\,Myr VLM stars and $\sim 500-1000$\,Myr brown dwarfs
are found to rotate rapidly, and yet to be substantially inactive. However, there
is evidence for a decrease in activity as VLM stars and brown dwarfs age, since 
[L$_{\rm H\alpha}$/L$_{\rm bol}$] falls from $\sim -4.5$-$-4.8$
to $\sim -5.1$-$-5.3$ as we move from Pleiades age brown dwarfs to $\sim 500-1000$\,Myr
brown dwarfs like LP\,944-20 and DENIS-P\,J1228.2-1547.

It should be noted however, that the possible identification of decreasing mass as a
parameter well correlated with the violation of the rotation-activity connection,
does not necessarily imply that mass is the controlling physical parameter.
The absence of chromospheric activity implies either the absence of magnetic fields, 
or the suppression of the chromospheric heating mechanism. In either case this could
be physically due to luminosity or effective temperature dependent effects (eg. the formation
of dust which begins at VLM effective temperatures), as well as to the internal structure
imposed on these objects by their mass.

Confirmation
of the rapid rotation of these objects via the search for periodic modulation
needs to be made a high priority. Many very low-mass stars show
periodic modulations due to star spots. Even if the inactive, rapid rotators
have no spots, they may exhibit periodic variation due to the transit
across their disk of dust clouds. In either case (spots or clouds) the result
will be the passage of regions of apparently low effective temperature
across their disk. Such
regions should show a clear signature in the regions of the spectrum
where TiO absorption is strong in late M-dwarfs, but absent in cooler objects like
DENIS-P\,J1228 (Tinney \etal\ 1997).

The inactive, rapid rotators represent a direct counter-example
to the generally held picture of the activity-rotation connection. 
Unravelling of this mystery will provide valuable insights into both the
interiors of brown dwarfs, and the mechanics of magnetic field and chromosphere
generation.

\section{Conclusion}

We have obtained high resolution optical spectroscopy of a sample of
VLM stars and one brown dwarf. Analysis of the kinematics of these stars
show no evidence for a significant kinematic difference between either
our photometrically-selected or proper-motion-selected sample,
and a sample of more massive nearby M-dwarfs. 

We present 0.43\AA\ resolution observations of two VLM stars and one 
intermediate age brown dwarf, for use as a spectroscopic atlas, together
with identifications for their major stellar atomic and molecular features.
Pseudo-equivalent widths for the Cs\,I and Rb\,I features present in these
are obtained. The Cs\,I\,8521\AA\ feature in particular shows promise
for discriminating L-type from M-type dwarfs in low resolution spectra.

Measurements of the H$\alpha$ emission equivalent widths and stellar
rotations are presented. These reveal that the brown dwarf LP\,944-20
is an ``inactive, rapid rotator'', like the M-dwarf BRI\,0021-0214
and the VLM Pleiades objects of Oppenheimer \etal\ (1997). This
discovery shows that violation of the rotation-activity connection
continues below the bottom of the H-burning main sequence. The fact
that such violation occurs for both Pleiades-age VLMs and
for a $\sim$500\,Myr old brown dwarf, indicates that age
is not the determining factor in such a violation.

\subsection*{Acknowledgments}

We would like to thank the technical and astronomical support staff at 
the ESO La Silla and Palomar Observatory for their most professional assistance throughout 
this observing program. The authors would like to thank X.Delfosse for communicating
the results of OHP observations prior to publication, and J.Mould for helpful discussion.

\label{lastpage}
\end{document}